\documentclass[12pt,fleqn]{article}
\pdfoutput=1
\usepackage[DIV12]{typearea}
\usepackage{amsmath,amsfonts,amssymb}
\usepackage[fleqn]{mathtools}
\usepackage{graphicx}
\usepackage{cite}
\usepackage{mathrsfs}
\usepackage{bbm}
\usepackage{pifont}
\usepackage{units}
\usepackage[small,loose]{subfigure}  
\usepackage{xcolor}
\usepackage{cancel}
\usepackage{pstricks,pst-node}

\usepackage[bookmarks=false]{hyperref}

\newcommand{\Eqref}[1]{equation~\eqref{#1}}
\newcommand{\Secref}[1]{section~\ref{#1}}
\newcommand{\Figref}[1]{figure~\ref{#1}}
\newcommand{\Tabref}[1]{table~\ref{#1}}

\newcommand{\eVdist}{\kern-0.06em}
\newcommand{\Ev}{\text{e\eVdist V}}     

\newcommand{\ev}{\:\text{e\eVdist V}}   

\newcommand{\gev}{\:\text{Ge\eVdist V}}
\newcommand{\tev}{\:\text{Te\eVdist V}}

\DeclareMathOperator{\re}{Re}
\DeclareMathOperator{\im}{Im}

\DeclareMathOperator{\diag}{diag}

\newcommand{\D}{\mathrm{d}}
\newcommand{\I}{\mathrm{i}}

\newcommand{\SU}[1]{\ensuremath{\mathrm{SU}(#1)}}

\newcommand{\Z}[1]{\ensuremath{\mathbbm{Z}_{#1}}} 
\newcommand{\A}[1]{\ensuremath{\mathrm{A}_{#1}}}

\newcommand{\AfourFlavon}{\ensuremath{\Phi}}
\newcommand{\AfourFlavonA}{\ensuremath{{\Phi_\nu}}}
\newcommand{\AfourFlavonB}{\ensuremath{{\Phi_e}}}

\newcommand{\rep}[1]{\ensuremath{\boldsymbol{#1}}}

\allowdisplaybreaks[1]

\numberwithin{equation}{section}
\numberwithin{table}{section}

\def\mytitle{Predictivity of models with spontaneously broken non--Abelian discrete flavor
symmetries}
\title{\mytitle}

\begin{document}

\begin{titlepage}

\begin{flushright}
TUM-HEP 877/13\\
UCI-TR-2013-03\\
FLAVOUR(267104)-ERC-36\\
CETUP*-12/018
\end{flushright}

\vspace*{1.0cm}

\begin{center}
{\Huge\bf
\mytitle
}

\vspace{1cm}

\textbf{Mu--Chun Chen\footnote[1]{Email: \texttt{muchunc@uci.edu}}{}}
\\[3mm]
\textit{\small
Department of Physics and Astronomy, University of California,\\
~~Irvine, California 92697--4575, USA
}
\\[5mm]
\textbf{
Maximilian Fallbacher\footnote[2]{Email:
\texttt{maximilian.fallbacher@ph.tum.de}}{},
Yuji Omura\footnote[3]{Email: \texttt{yuji.omura@tum.de}}{},
Michael Ratz\footnote[4]{Email: \texttt{michael.ratz@tum.de}}{},
Christian Staudt\footnote[5]{Email: \texttt{christian.staudt@ph.tum.de}}{}
}
\\[3mm]
\textit{\small
Physik Department T30, Technische Universit\"at M\"unchen, \\
~~James--Franck--Stra\ss e, 85748 Garching, Germany
}
\end{center}

\vspace{1cm}

\begin{abstract}
In a class of supersymmetric flavor models predictions are based on residual
symmetries of some subsectors of the theory such as those of the charged leptons
and neutrinos. However, the vacuum expectation values of the so--called flavon
fields generally modify the K\"ahler potential of the setting, thus changing the
predictions. We derive simple analytic formulae that allow us to understand the
impact of these corrections on the predictions for the masses and mixing
parameters. Furthermore, we discuss the effects on the vacuum alignment and on
flavor changing neutral currents. Our results can also be applied to 
non--supersymmetric flavor models.
\end{abstract}

\end{titlepage}

\section{Introduction}

\begin{figure}[!h!]
  \centering
  \includegraphics{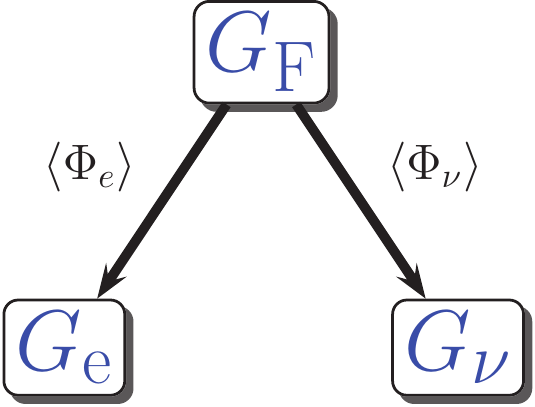}
  \caption{The flavor symmetry $G_\mathrm{F}$ gets broken to different subgroups in
   different sectors of the theory.}
\label{fig:VEVmisalignment}
\end{figure}

Explanations of the observed pattern of fermion masses and mixing are often
based on spontaneously broken flavor symmetries. In this paper, we concentrate
on supersymmetric models attempting to explain the observed flavor structure by
discrete symmetries. At some (high) energy scale, the flavor symmetry, denoted
by $G_\mathrm{F}$ in what follows, is spontaneously broken by some appropriate
`flavon' fields, which acquire vacuum expectation values (VEVs). Although our
analysis also applies to non--supersymmetric settings,  we base our discussion
on the  lepton sector of supersymmetric extensions of the standard model (SM). 
In order to be specific, consider a prototype superpotential of the form
\begin{equation}\label{eq:leadingW}
 \mathscr{W}_\mathrm{leading}
 ~=~
 \frac{1}{\Lambda}(\AfourFlavonB)_{gf}\,L^g\,R^f\,H_d
 +\frac{1}{\Lambda\,\Lambda_\nu}(\AfourFlavonA)_{gf}\,L^g\,H_u\,L^f\,H_u\;,
\end{equation}                                                             
where $L^g$ and $R^f$ (with the flavor indices $1\le f,g\le 3$) denote the
lepton doublets and singlets, respectively, whereas $H_u$ and $H_d$ are the usual Higgs
doublets of the supersymmetric standard model. The two scales involved are the cut--off
scale $\Lambda$ of the theory and the see--saw scale $\Lambda_\nu$. For the sake of
definiteness, we shall take $\Lambda$ to be around the unification scale, although
our results will not depend on this choice.  $\Phi_e$ and $\Phi_\nu$ denote the
flavons, which acquire VEVs that are assumed to be somewhat below $\Lambda$
such that the expansion parameters of our theory are
$\langle\Phi_e\rangle/\Lambda$ and $\langle\Phi_\nu\rangle/\Lambda$. Inserting
the flavon VEVs leads to an effective superpotential
\begin{equation}\label{eq:Weff}
 \mathscr{W}_\mathrm{eff}~=~(Y_e)_{gf}\,L^g\,R^f\,H_d
 +\frac{1}{4}\kappa_{gf}\,L^g\,H_u\,L^f\,H_u\;.
\end{equation}
One is often left with a situation in which neither $\Phi_e$ nor $\Phi_\nu$
breaks $G_\mathrm{F}$ completely, but respect the residual symmetries
$G_\mathrm{e}$ and $G_\nu$, respectively (cf.\
figure~\ref{fig:VEVmisalignment}), while the intersection of the residual
symmetries is smaller or empty. Given that higher--order terms are either 
subleading or may be completely forbidden by some appropriate symmetries such
as $R$ symmetries, these residual symmetries allow us to make predictions.

Models that make predictions based on such residual symmetries have become
rather popular in the past (see e.g.~\cite{Altarelli:2010gt,King:2013eh}). One
can, for example, successfully obtain the bi--maximal mixing
pattern~\cite{Vissani:1997pa,Barger:1998ta} as well as the tri--bi--maximal (TBM)
mixing pattern~\cite{Harrison:2002er,Ma:2004zv}. 

The potential problem with such predictions is that they are based on the
holomorphic superpotential only. However, there are modifications coming from
the K\"ahler potential \cite{Leurer:1992wg, Leurer:1993gy,Dudas:1995yu}. Given
the fact that for most of the proposed patterns the mixing parameters, i.e.\
mixing angles and phases, run under the renormalization group (RG), and that in
supersymmetric theories RG corrections affect the K\"ahler potential only, it is
clear that it will be nearly impossible to avoid such corrections. One may,
therefore, question how solid the predictions based only on the holomorphic
sector really are.  Clearly, the canonical K\"ahler potential does not include
all terms allowed by the flavor symmetry. Rather, if one is to derive
predictions from higher--order terms in the expansion parameters
$\langle\Phi_{e/\nu}\rangle/\Lambda$, one should take into account both the
superpotential and the K\"ahler potential. The full K\"ahler potential is
\begin{equation}
 K~=~K_\mathrm{canonical}+\Delta K\;,
\end{equation}
where the canonical part is given by (only considering the leptons)
\begin{equation}
 K_\mathrm{canonical}~\supset~
 \left(L^f\right)^\dagger\,\delta_{fg}\,L^g
 +\left(R^f\right)^\dagger\,\delta_{fg}\,R^g
 \;.
\label{eq:Kcanonical}
\end{equation}
$\Delta K$ includes contractions of $L^f$ and $R^f$ with the flavons,
such as $(L\Phi_{e/\nu})^\dagger(L\Phi_{e/\nu})$, which may not be forbidden by
any (conventional) symmetry, and it has the general form
\begin{equation}
  \Delta K~=~
  \left(L^f\right)^\dagger \, (\Delta \mathscr{K}_L)_{fg} \,L^g 
  + \left(R^f\right)^\dagger \, (\Delta \mathscr{K}_R)_{fg} \,R^g
  \;.
\end{equation}
Here $\Delta \mathscr{K}_{L}$ and $\Delta \mathscr{K}_{R}$ are Hermitean
matrices which describe the modification of the K\"ahler metric after the
flavons acquire their VEVs. The structure of these Hermitean matrices,
therefore, depends on the flavor group and the flavon VEVs.  

After the breaking of the flavor symmetry, one needs to redefine the fields
in order to return to a canonical K\"ahler potential~\cite{Soni:1983rm,King:2003xq,Espinosa:2004ya}. As
we shall discuss in more detail below, in this new basis generically none of
the subsectors exhibits a residual symmetry. Among other things, this explains
why the parameters run even though their values appear to be determined by
$G_\mathrm{e}$ and $G_\nu$, respectively. The crucial property of $\Delta K$ is that its size will, in general, be
controlled by the above expansion parameters $\langle\Phi_e\rangle/\Lambda$ and
$\langle\Phi_\nu\rangle/\Lambda$, i.e.\ the very same quantities that set the
scale of the entries of the mass and coupling matrices in the effective
superpotential $\mathscr{W}_\mathrm{eff}$. In addition, $\Delta K$ will depend
on K\"ahler coefficients which multiply the above contractions and are hard to
determine in an effective field theory approach.

Using methods previously used for the renormalization group equations (RGEs)
in~\cite{Antusch:2003kp,Antusch:2005gp}, one can obtain an analytic
understanding of the K\"ahler corrections~\cite{Antusch:2007ib,Antusch:2007vw,Chen:2012ha}. As we pointed out
in~\cite{Chen:2012ha}, the corresponding corrections are sizable and will in
general lead to a strong modification of the predictions. In particular, they
may render patterns that appeared to be ruled out, such as the TBM one,
consistent with observation --- and vice versa.

The purpose of this paper is to extend our discussion of these changes by
presenting a full derivation of the analytic formulae.  We start out in
\Secref{sec:HolomorphicTerms} by reviewing the predictions from the
superpotential of two well--known models, one of which is based on
\A4~\cite{Altarelli:2005yx} and the other on
$\mathrm{T}'$~\cite{Chen:2009gf}, and also compare the results to the current
best fit values. In \Secref{sec:Kaehler} we provide an analytic discussion of
the K\"ahler corrections. We then apply our analytic understanding 
to the two sample models in
\Secref{sec:implications}, in which we also comment on the
implications of K\"ahler corrections for the VEV alignment and for flavor
changing neutral currents (FCNCs). Finally, \Secref{sec:conclusions} summarizes 
our results.

\section{Mixing parameters from the superpotential}
\label{sec:HolomorphicTerms}

In this section we review by means of two simple examples how predictions based
on residual symmetries of the mass terms in the superpotential are derived.

\subsection{TBM from \boldmath \A4 \unboldmath}
\label{sec:A4Model}

One of the simplest and most popular choices of a flavor symmetry group is
\A4~\cite{Ma:2004zv}. The resulting mixing is characterized by the TBM mixing
matrix
\begin{equation}
\label{eq:TBMmatrix}
 U_\mathrm{TBM} ~=~ \begin{pmatrix}
                \sqrt{\frac{2}{3}} & \frac{1}{\sqrt{3}} & 0 \\
		- \frac{1}{\sqrt{6}} & \frac{1}{\sqrt{3}} & - \frac{1}{\sqrt{2}} \\
	      - \frac{1}{\sqrt{6}} & \frac{1}{\sqrt{3}} &  \frac{1}{\sqrt{2}}
	      \end{pmatrix}\;,
\end{equation}
which leads to the mixing angles in standard parametrization (cf.\
appendix~\ref{app:parametrization}) shown in \Tabref{tab:angles}.
\begin{table}[t]
\centering
\renewcommand{\arraystretch}{1.5}
\begin{tabular}{lccc} \hline
 & $\theta_{12}$ & $\theta_{13}$ & $\theta_{23}$\\ \hline
 TBM prediction: & $\arctan{\left( \sqrt{0.5} \right)} \approx 35.3^\circ$ &  $0$ & $45^\circ$\\
 Best fit values $(\pm 1 \sigma)$: & 
 $\left(33.6^{+1.1}_{-1.0}\right)^\circ$ & 
 $\left(8.93^{+0.46}_{-0.48}\right)^\circ$ & 
 $\left(38.4^{+1.4}_{-1.2}\right)^\circ$\\ \hline
\end{tabular}
\caption{Tri--bi--maximal prediction for the neutrino mixing angles and best fit
values from the global fit by \cite{Fogli:2012ua}.}
\label{tab:angles}
\end{table}
The measurement of $\theta_{13}$ \cite{Abe:2011fz,An:2012eh,Ahn:2012nd} revealed
a considerable deviation from the tri--bi--maximal prediction and also the
recent best fit values from global analyses for $\theta_{23}$ are in tension with maximal mixing. 
Yet, the TBM pattern may still serve as a good first order approximation of
the observed mixing angles.

As common to many flavor models, the three generations of left--handed lepton
doublets are assumed to transform as a triplet under \A4,  $L \sim$ \rep{3}. The
three singlet representations of \A4, \rep{1}, \rep{1''} and \rep{1'}, are
assigned to the right--handed charged leptons $e_\mathrm{R}$, $\mu_\mathrm{R}$
and $\tau_\mathrm{R}$, respectively, and the Higgs fields $H_{u}$ and $H_{d}$
are not charged under the flavor symmetry. The mass matrices are generated by
VEVs of three flavon fields: the two \A4 triplets \AfourFlavonA\ and
\AfourFlavonB, and the pure singlet $\xi \sim$ \rep{1}. At leading order in the
ratio flavon VEV over the cut--off scale, the terms leading to the Yukawa couplings
and to the Weinberg neutrino operator (cf.\ \Eqref{eq:leadingW}) read 
\begin{eqnarray}
 \mathscr{W}_\nu
 & = &
 \frac{\lambda_1}{\Lambda\,\Lambda_{\mathrm{\nu}}}\,\left\{\left[(L\,H_u) \otimes
(L\,H_u)\right]_{\rep{3}_\mathrm{s}} \otimes
\AfourFlavonA\right\}_{\rep{1}}
 +
 \frac{\lambda_2}{\Lambda\,\Lambda_{\mathrm{\nu}}}\,\left[(L\,H_u) \otimes (L\,H_u)\right]_{\rep{1}}\,\xi\;,
 \\
 \mathscr{W}_e
 & = &
\frac{h_e}{\Lambda}\,\left(\AfourFlavonB\otimes L\right)_{\rep{1}}\,H_d\,e_\mathrm{R}
 +
 \frac{h_\mu}{\Lambda}\,\left(\AfourFlavonB\otimes L\right)_{\rep{1'}}\,H_d\,\mu_\mathrm{R}
 +
 \frac{h_\tau}{\Lambda}\,\left(\AfourFlavonB\otimes L\right)_{\rep{1''}}\,H_d\,\tau_\mathrm{R}
 \;,
\end{eqnarray}
where again $\Lambda$ and $\Lambda_{\nu}$ denote the cut--off and the see--saw scale,
respectively.

In order to distinguish the flavon field \AfourFlavonA, which couples to the neutrinos,   
from the flavon field \AfourFlavonB, which couples to the charged leptons, one introduces an additional \Z4 symmetry. Under this
symmetry, \AfourFlavonA\ changes sign whereas \AfourFlavonB\ stays invariant.
Furthermore, under the  \Z4  symmetry, $\xi \rightarrow - \xi$, $L \rightarrow \I L$ and $R
\rightarrow - \I R$.

The desired tri--bi--maximal lepton mixing is achieved when the flavons
acquire VEVs in the directions
\begin{subequations}\label{eq:A4VEVpattern}
\begin{eqnarray}
\langle\AfourFlavonA\rangle & = & \left(v,v,v\right)\;, \\
\langle\AfourFlavonB\rangle & = & \left(v',0,0\right)\;, \\
\langle\xi\rangle & = & w\;.
\end{eqnarray}
\end{subequations}
This choice breaks the flavor symmetry $G_\mathrm{F}=\A4\times\Z4$ to
$G_\mathrm{e}=\Z3 \times \Z4$ and $G_\nu=\Z2 \times \Z2$ in the charged lepton
and neutrino sector, respectively. These residual symmetries lead to TBM. This
can be seen explicitly by computing the mass matrices after electroweak
symmetry breaking. The charged lepton mass matrix reads
\begin{equation}
 m_{e} ~=~ v_{d}\, \diag\left(y_{e},\, y_{\mu},\, y_{\tau}\right)\;,
\end{equation}
where $v_{d}$ is the VEV of the down--type Higgs and
$y_{e,\,\mu,\,\tau}=h_{e,\,\mu,\,\tau}\,\frac{v'}{\Lambda}$. Here and in
the following we work in a basis in which the charged lepton Yukawa matrix is
diagonal.

On the other hand, in this basis the neutrino mass matrix is
non--diagonal. Using the abbreviations  $a=2
\lambda_{2}\, \frac{v_{u}^{2}}{\Lambda_{\mathrm{\nu}}}\,\frac{w}{\Lambda}$ and 
$d= \sqrt{2}  \lambda_{1}\, \frac{v_{u}^{2}}{\Lambda_{\nu}}\,\frac{v}{\Lambda}$,
where $v_{u}$ is the VEV of the up--type Higgs, it can be written as
\begin{equation}\label{eq:mnu}
 m_{\nu} ~=~ 
 \begin{pmatrix}
                a + 2d & -d & -d \\
		-d & 2d & a -d \\
                -d & a - d & 2d
\end{pmatrix}\;.
\end{equation}
The lepton mixing matrix $U_\mathrm{PMNS}$ is then the unitary transformation
that diagonalizes the neutrino mass matrix and is indeed given by the
tri--bi--maximal matrix \eqref{eq:TBMmatrix}.

Even though this simple model seems to be excluded by the recent measurements,
there are still many loopholes which can make it viable. For example, there are
attempts to explain the deviations by higher--order terms in the superpotential,
cf.\ \cite{Altarelli:2012ss} and references therein. However, as we have shown
in \cite{Chen:2012ha}, there are also corrections due to higher--order K\"ahler
potential terms, which can either reconcile the model predictions with data or
drive them even further away.

\subsection{\boldmath $\mathrm{T}'$ \unboldmath}
\label{sec:tprime}

Another interesting example model is based on the double covering group of \A4,
which is called $\mathrm{T}'$. Like \A4, this group contains three irreducible
singlet representations and one triplet. Additionally, the group contains three
doublet representations \rep{2}, \rep{2'} and \rep{2''}. The specific model
\cite{Chen:2009gf} we discuss comes with several flavon fields, which are
summarized in \Tabref{tab:tprime}, and also two additional Abelian \Z{12}
symmetries.
\begin{table}[!h]
\centering
\begin{tabular}{lcccccccc} \hline
 & $\phi$ & $\phi'$ & $\psi$ & $\psi'$ & $\zeta$ & $N$ & $\xi$ & $\eta$  \\ \hline
$\mathrm{T}'$ & \rep{3} &  \rep{3} & \rep{2'} & \rep{2} & \rep{1''} & \rep{1'} & \rep{3} & \rep{1} \\
\Z{12} & 3 &  2 & 6 & 9 & 9 & 3 & 10 & 10 \\
\Z{12} & 3 &  6 & 7 & 8 & 2 & 11 & 0 & 0 \\ \hline
\end{tabular}
\caption{Flavon content of the $\mathrm{T}'$ model.}
\label{tab:tprime}
\end{table}
The flavons acquire VEVs along the directions
\begin{eqnarray}
 \langle \phi \rangle & =& \phi_0\, 
 \left(\begin{array}{c}1\\0\\0\end{array}\right)\;, \quad  
 \langle \phi' \rangle~=~ 
 \phi'_0 \left(\begin{array}{c}1\\1\\1\end{array}\right)\;,\quad  
\langle \xi \rangle~ =~ \xi_0\, 
\left(\begin{array}{c}1\\1\\1\end{array}\right)\;, \nonumber \\
 \langle \psi \rangle & = & \psi_0\, 
 	\left(\begin{array}{c}1\\0\end{array}\right)\;,\quad  
	\langle \psi' \rangle ~=~ \psi'_0\, 
	\left(\begin{array}{c}1\\1\end{array}\right)\;, 
\label{eq:tprimeVEV}
\end{eqnarray}
and the fields transforming as one--dimensional representations, $\zeta$,
$N$ and $\eta$, assume non--trivial values. With this choice of VEVs, the model \cite{Chen:2009gf} gives rise to near tri--bi--maximal lepton
mixing, 
\begin{equation}
\theta_{12}~\approx~33^\circ\;, \quad \theta_{23}~=~45^\circ
\quad \text{and} \quad \theta_{13}~\approx~3^\circ\;.
\label{eq:TprimeStart}
\end{equation}
The deviations from the exact TBM mixing pattern are due to the corrections from the charged lepton sector, 
and they are related to the Cabibbo angle through the $\SU5$ GUT relations.  
Furthermore, the model also predicts a leptonic Dirac CP violating phase from the
superpotential and an absolute neutrino mass scale, e.g.\ $m_1=0.0156\,\Ev$ for
mass--squared differences given by $\Delta m_{21}^2 ~=~ 8.0 \cdot
10^{-5}\,(\Ev)^2$ and $\Delta m_{32}^2 ~=~ 2.4 \cdot 10^{-3}\,(\Ev)^2$.

\section{Corrections due to K\"ahler potential terms}
\label{sec:Kaehler}

Let us now look at the K\"ahler potential of the theory. As already
mentioned, higher--order terms will, after the flavons acquire their VEVs, lead
to a non--canonical K\"ahler metric. Let us spell this out in more detail, using
the \A4 and the $\mathrm{T}'$ examples from \Secref{sec:HolomorphicTerms}.
Here, the left--handed lepton doublets transform as a triplet of either \A4 or
$\mathrm{T}'$, respectively. Contractions of these triplets with the flavons
will then lead to a K\"ahler metric with off--diagonal terms after the
flavons acquire a VEV.

\subsection{Linear flavon corrections from left--handed leptons}
\label{sec:linear}

We start with terms which are linear in the flavons. Focussing on the \A4 
symmetry only, these linear contributions are given by
\begin{equation}
 \Delta K_\mathrm{linear} ~=~
 \sum\limits_{i\, \in \{\mathrm{a},\mathrm{s}\}}
 \left(\frac{\kappa_{\AfourFlavonA}^{(i)}}{\Lambda}\,
 {L^\dagger\,(L \otimes \AfourFlavonA)_{\rep{3}_i}}
 + \frac{\kappa_{\AfourFlavonB}^{(i)}}{\Lambda}\,
 {L^\dagger\,(L \otimes \AfourFlavonB)_{\rep{3}_i}}\right)
 + \frac{\kappa_\xi}{\Lambda} \,
 {\xi \, L^\dagger L}
 +\text{h.c.}\;,
\end{equation}
and are suppressed by only one power of the expansion parameter 
$\langle\Phi_{e/\nu}\rangle/\Lambda$. The last term does not lead to a change of
the mixing parameters because it just changes the overall normalization of the
kinetic term of the lepton doublets. On the other hand, the terms containing
\AfourFlavonA\ and \AfourFlavonB\ do modify the model predictions. The
contractions with a generic triplet flavon \AfourFlavon\ are
\begin{subequations}
\begin{eqnarray}
{L^\dagger\,(L \otimes \AfourFlavon)_{\rep{3}_\mathrm{s}}}
 & = & 
 \frac{1}{\sqrt{2}}\,\left[(L_{1}^\dagger)(2\,L_{1} \AfourFlavon_1 - L_{2} \AfourFlavon_3 - L_{3} \AfourFlavon_2)\right. 
 +
 (L_{2}^\dagger)(2\,L_{3} \AfourFlavon_3 - L_{1} \AfourFlavon_2 - L_{2} \AfourFlavon_1) \nonumber \\
 &  &{}\hphantom{\frac{1}{\sqrt{2}}\,\left[\right.}+ 
 \left. (L_{3}^\dagger)(2\,L_{2} \AfourFlavon_2 - L_{1} \AfourFlavon_3 - L_{3} \AfourFlavon_1)\right]\;, \\
{L^\dagger\,(L \otimes \AfourFlavon)_{\rep{3}_\mathrm{a}}}
 & = & 
 \I\, \sqrt{\frac{3}{2}}\,\left[(L_{1}^\dagger)(L_{2} \AfourFlavon_3 - L_{3} \AfourFlavon_2)\right. \nonumber
 +
 (L_{2}^\dagger)(L_{1} \AfourFlavon_2 - L_{2} \AfourFlavon_1) \nonumber \\
 &  &{}\hphantom{\I\, \sqrt{\frac{3}{2}}\,\left[\right.}+ 
 \left. (L_{3}^\dagger)(L_{3} \AfourFlavon_1 - L_{1} \AfourFlavon_3)\right]\;.
\end{eqnarray}
\end{subequations}
Plugging in the flavon VEVs leads to departures from the canonical K\"ahler
metric,
\begin{equation}
  \mathscr{K}_{fg}
  ~=~
  \left(\frac{\partial^2 K}{\partial (L^f)^\dagger\,\partial L^g}\right)
  ~=~
  \delta_{fg}
  ~\xrightarrow{\AfourFlavon\to\langle\AfourFlavon\rangle}~
  \delta_{fg}+(\Delta \mathscr{K})_{fg}\;.
\end{equation}
In what follows, we will find it convenient to decompose $\Delta \mathscr{K}$
according to
\begin{equation}
 (\Delta \mathscr{K})_{fg}
 ~=~
 \alpha\,P_{fg}+\text{h.c.}\;,
\end{equation}
where $P$ encodes the matrix structure and $\alpha$ is a continuous parameter
reflecting the size of the K\"ahler correction. For the flavon VEV
$\langle\AfourFlavon\rangle=\langle\AfourFlavonB\rangle=(v',0,0)$  one obtains
the K\"ahler corrections
\begin{subequations}
\begin{eqnarray}
 (\Delta \mathscr{K})_{\AfourFlavonB}^{(\mathrm{s})} 
 & = & 
 \kappa_{\AfourFlavonB}^{(s)}\, v' \, \frac{1}{\sqrt{2}}\,P_{\AfourFlavonB}^{(\mathrm{s})}+\text{h.c.}\;,\\
 (\Delta \mathscr{K})_{\AfourFlavonB}^{(\mathrm{a})} & = &  \I\, \kappa_{\AfourFlavonB}^{(a)}\, v'\, \sqrt{\frac{3}{2}}\,
 P_{\AfourFlavonB}^{(\mathrm{a})} +\text{h.c.} 
\end{eqnarray}
\end{subequations}
with the $P$ matrices
\begin{subequations}
\begin{eqnarray}
\label{eq:linP1}
P_{\AfourFlavonB}^{(\mathrm{s})} & = & \diag(2,-1,-1)
\;,\\
\label{eq:linP2}
P_{\AfourFlavonB}^{(\mathrm{a})} & = & \diag(0,-1,1)
\;,
\end{eqnarray}
\end{subequations}
whereas for $\langle\AfourFlavon\rangle=\langle\AfourFlavonA\rangle=(v,v,v)$
one gets
\begin{subequations}
\begin{eqnarray}
 (\Delta \mathscr{K})_{\AfourFlavonA}^{(\mathrm{s})} & = & \kappa_{\AfourFlavonA}^{(s)}\, v \, \frac{1}{\sqrt{2}}\,
 P_{\AfourFlavonA}^{(\mathrm{s})}+\text{h.c.}\;,\\
(\Delta \mathscr{K})_{\AfourFlavonA}^{(\mathrm{a})} & = & \I\, \kappa_{\AfourFlavonA}^{(a)}\, v\, \sqrt{\frac{3}{2}}\,
P_{\AfourFlavonA}^{(\mathrm{a})}+\text{h.c.}
\end{eqnarray}
\end{subequations}
with
\begin{subequations}
\begin{eqnarray}
P_{\AfourFlavonA}^{(\mathrm{s})} & = & 
\left(
\begin{array}{ccc}
 2 & -1 & -1 \\
 -1 & -1 & 2 \\
 -1 & 2 & -1
\end{array}
\right)\;,\\
P_{\AfourFlavonA}^{(\mathrm{a})} & = & 
\left(
\begin{array}{ccc}
 0 & 1 & -1 \\
 1 & -1 & 0 \\
 -1 & 0 & 1
\end{array}
\right)\;.
\end{eqnarray}
\end{subequations}

However, in the specific \A4 model from \Secref{sec:A4Model}, the terms
comprising \AfourFlavonA\ are forbidden by the additional \Z4 symmetry.
Therefore, we only get corrections in this case from the matrices in
\Eqref{eq:linP1} and \Eqref{eq:linP2}. One may introduce additional symmetries
in such a way that all flavons are charged (like in the $\mathrm{T}'$
example in \Secref{sec:tprime}), and hence forbid linear flavon
contributions in the K\"ahler potential all together. Therefore, these linear corrections
will not necessarily spoil the predictivity of a given model. This is different
in the case of quadratic corrections, which we discuss next.

\subsection{Second order corrections from left--handed leptons}
\label{sec:quadratic}

Unlike the linear terms, some of the quadratic corrections to the K\"ahler
potential, which are of the form  $(L \otimes \Phi_i)^\dagger (L \otimes
\Phi_j)$, cannot be forbidden by any (conventional) symmetry. Obviously, terms
with $i\neq j$ can again be forbidden by a symmetry, however, terms like $(L
\otimes \AfourFlavon)^\dagger (L \otimes \AfourFlavon)$ with
$\AfourFlavon=\AfourFlavonA$ or $\AfourFlavonB$ cannot. We will comment later in
\Secref{sec:conclusions} how one may control or avoid such corrections in more
complete settings.  In the \A4 model we get six different possible terms for
each flavon,  $(L \otimes \AfourFlavonA)_{\rep{R}}^\dagger (L \otimes
\AfourFlavonA)_{\rep{R'}}$ and $(L \otimes \AfourFlavonB)_{\rep{R}}^\dagger (L
\otimes \AfourFlavonB)_{\rep{R'}}$, e.g.\  $(L \otimes
\AfourFlavonA)_{\rep{3}_s}^\dagger (L \otimes \AfourFlavonA)_{\rep{3}_a}$. Using
the \A4 multiplication rules from appendix~\ref{app:A4}, the latter term can be
recast as
\begin{eqnarray}
\lefteqn{
(L \otimes \AfourFlavonA)^{\dagger}_{\rep{3}_s}(L \otimes 
\AfourFlavonA)_{\rep{3}_a}}\nonumber\\
 & = & 
 \I\, \frac{\sqrt{3}}{2}\, \left[\left(2\,L_1^\dagger \AfourFlavonA_1^\dagger - L_2^\dagger  \AfourFlavonA_3^\dagger -L_3^\dagger  \AfourFlavonA_2^\dagger\right)\left(L_2  \AfourFlavonA_3 - L_3  \AfourFlavonA_2\right)\right. \nonumber \\
 &&{}\hphantom{\I\, \frac{\sqrt{3}}{2}\, \left[\right.} 
 +\left(2L_3^\dagger\AfourFlavonA_3^\dagger - L_2^\dagger\AfourFlavonA_1^\dagger
 -L_1^\dagger\AfourFlavonA_2^\dagger\right)\left(L_1\AfourFlavonA_2 -
 L_2\AfourFlavonA_1\right) \nonumber\\
 &&{}  \hphantom{\I\, \frac{\sqrt{3}}{2}\, \left[\right.} 
 +\left.\left(2L_2^\dagger\AfourFlavonA_2^\dagger - L_1^\dagger\AfourFlavonA_3^\dagger -L_3^\dagger\AfourFlavonA_1^\dagger\right)\left(L_3\AfourFlavonA_1 - L_1\AfourFlavonA_3\right)\right]\;
\end{eqnarray}
and leads to the K\"ahler correction
\begin{equation}
\Delta \mathscr{K} ~=~\I\, \kappa\,\frac{\sqrt{3}}{2}\,P+ \text{h.c.}
\end{equation}
with the $P$ matrix
\begin{equation}
P ~=~ 
\begin{pmatrix} 
-\AfourFlavonA_2^\dagger\AfourFlavonA_2 + \AfourFlavonA_3^\dagger\AfourFlavonA_3 & 2\AfourFlavonA_1^\dagger\AfourFlavonA_3+\AfourFlavonA_2^\dagger\AfourFlavonA_1 & -2\AfourFlavonA_1^\dagger\AfourFlavonA_2-\AfourFlavonA_3^\dagger\AfourFlavonA_1\\
-2\AfourFlavonA_2^\dagger\AfourFlavonA_3-\AfourFlavonA_1^\dagger\AfourFlavonA_2 & -\AfourFlavonA_3^\dagger\AfourFlavonA_3 + \AfourFlavonA_1^\dagger\AfourFlavonA_1 & 2\AfourFlavonA_2^\dagger\AfourFlavonA_1+\AfourFlavonA_3^\dagger\AfourFlavonA_2\\
2\AfourFlavonA_3^\dagger\AfourFlavonA_2+\AfourFlavonA_1^\dagger\AfourFlavonA_3 & -2\AfourFlavonA_3^\dagger\AfourFlavonA_1-\AfourFlavonA_2^\dagger\AfourFlavonA_3 & \AfourFlavonA_2^\dagger\AfourFlavonA_2 - \AfourFlavonA_1^\dagger\AfourFlavonA_1
\end{pmatrix}.
\label{eq:PVderivation}
\end{equation}
We then get for $\langle\AfourFlavonA\rangle=(v,v,v)$
\begin{equation}
 \Delta \mathscr{K}~=~\I\,\frac{\sqrt{3}}{2}\, \kappa\,v^2
\left(\begin{array}{ccc}    
 0 &  3 & -3\\
-3 &  0 &  3\\
 3 & -3 &  0
\end{array}\right) + \text{h.c.}
~=~3 \frac{\sqrt{3}}{2}\,\kappa\, v^2\, P_\mathrm{V} + \text{h.c.}\;.
\end{equation}

If we treat all of the possible twelve terms (six for each flavon) in such a
way, we get several corrections which lead to identical $P$ matrices. In total,
we can summarize them by five different matrices $P_{\mathrm{I}-\mathrm{V}}$,
where the derivation of the fifth matrix $P_\mathrm{V}$ has just been shown
above. The first three matrices
\begin{subequations}
\label{eq:KaehlerBasis}
\begin{equation}
 P_\mathrm{I}~=~\diag(1,0,0)\;,\quad
 P_\mathrm{II}~=~\diag(0,1,0)
 \quad\text{and}\quad
 P_\mathrm{III}~=~\diag(0,0,1)
\label{eq:pmatrix1}
\end{equation}
come from contractions of $L$ with $\AfourFlavonB$. Since
$\langle\AfourFlavonB\rangle=(v',0,0)$, their contribution in the K\"ahler
potential is proportional to $(v')^2$. The other two matrices,
\begin{equation}
 P_\mathrm{IV}~=~\begin{pmatrix}
 1 & 1 & 1 \\ 1 & 1 & 1\\ 1 & 1 & 1
 \end{pmatrix}
 \quad\text{and}\quad
 P_\mathrm{V}~=~\begin{pmatrix}
 0 & \I & -\I \\ -\I & 0 & \I\\ \I & -\I & 0
 \end{pmatrix}
 \;,
\label{eq:pmatrix2}
\end{equation}
\end{subequations}
are contributions due to $\AfourFlavonA$; therefore, their contribution is
proportional to $v^2$ since $\langle\AfourFlavonA\rangle=(v,v,v)$.
An important property of all these corrections is that they are
controlled by the square of our expansion parameters VEV over the fundamental scale
as well as some unknown coefficient in the K\"ahler potential.

\subsection{Corrections from the right--handed leptons}

As was already stated in the introduction one can also have corrections for
the right--handed lepton fields, depending on their representation under the
flavor group. In the \A4 example, the right--handed leptons are \A4
singlets, \rep{1}, \rep{1'} and \rep{1''}; therefore, corrections with the
flavon triplets are going to be diagonal, e.g.\ 
\begin{equation}
  K~\supset~ 
  \frac{1}{\Lambda^2}\,\left(e_\mathrm{R}\,\AfourFlavonA\right)^\dagger 
  \left(e_\mathrm{R}\,\AfourFlavonA\right) 
  ~=~ 
  e_\mathrm{R}^\dagger\,e_\mathrm{R}\,
  \frac{\AfourFlavonA^\dagger\,\AfourFlavonA}{\Lambda^2}\;,
\end{equation}
which, after VEV insertion, gives
$3\,|v|^2\,e_\mathrm{R}^\dagger\,e_\mathrm{R}/\Lambda^2$. We get similar terms
for $\mu_\mathrm{R}$, $\tau_\mathrm{R}$ and also for contractions of the
right--handed leptons with the flavon field $\AfourFlavonB$. The only other
flavon in the model is the \A4 singlet $\xi$; hence, we also have diagonal
corrections proportional to $|\langle \xi \rangle|^2$.

The model could also contain flavons which are in the singlet representation
\rep{1'} or \rep{1''}, in this case, non--diagonal corrections are possible due
to terms with two different flavons. However, these corrections, just like the
linear ones, can easily be forbidden by an additional symmetry. Therefore, we
focus on corrections for the right--handed lepton fields which are
diagonal, i.e.\ $P_R = \diag(\alpha_1,\alpha_2,\alpha_3)$ where the $\alpha_i$ are not related and
depend on the VEVs and arbitrary K\"ahler coefficients.
Since we are working in a basis with diagonal charged lepton Yukawa matrices, a
diagonal redefinition of the right--handed fields can only affect the mass
eigenvalues but not the mixing angles. This is also reflected by our
analytical formulae.

\subsection{Second order corrections for a model based on \boldmath $\mathrm{T}'$ \unboldmath}
\label{sec:tprimeQu}

We now extend our previous analysis to a model which is based on
$\mathrm{T}'$~\cite{Chen:2009gf}, the double covering group of \A4.  As stated
in \Secref{sec:tprime}, $\mathrm{T}'$, like \A4, contains three irreducible
one--dimensional representations and one triplet. Beyond this, the
multiplication law for the contraction of two triplets is the same. For doublets
and triplets, it is given by
\begin{equation}
 \rep{2},\rep{2'},\rep{2''} \otimes \rep{3} ~=~ \rep{2} \oplus \rep{2'} \oplus \rep{2''}\;.
\label{eq:TprimMult}
\end{equation}
A complete list of tensor products can be found, for instance, in
\cite{Ishimori:2010au}.  A closer look at \Tabref{tab:tprime} shows that in this
model there are no linear corrections in the K\"ahler potential, since all
flavons are charged under one of the \Z{12} symmetries. Nonetheless, there are
several second order corrections: looking at the VEV structure of the flavon
fields in \Eqref{eq:tprimeVEV}, it is obvious that we have the same K\"ahler
corrections as in the \A4 example due to the flavons $\phi$, $\phi'$ and $\xi$.
However, there are additional terms due to the doublet fields $\psi$ and
$\psi'$ through higher--order terms of the form
$\left(L\otimes\psi\right)_{\rep{R}}^\dagger\left(L\otimes\psi\right)_{\rep{R}}$
and
$\left(L\otimes\psi'\right)_{\rep{R}}^\dagger\left(L\otimes\psi'\right)_{\rep{R}}$.
The contributions of these terms can be determined with the multiplication law
in~\Eqref{eq:TprimMult}, which shows that \rep{R} can only be one of the three
doublets. Using this multiplication rule, we get, e.g., the contribution
\begin{eqnarray}
 \left(L \otimes \psi'\right)_{\rep{2'}}^\dagger\left(L \otimes \psi'\right)_{\rep{2'}} & = & \left(\sqrt{2}\,\psi'_2\,L_3 + \psi'_1\,L_2\right)^\dagger \left(\sqrt{2}\,\psi'_2\,L_3 + \psi'_1\,L_2\right) + \nonumber \\
& + & \left(\sqrt{2}\,\psi'_1\,L_1 - \psi'_2\,L_2\right)^\dagger \left(\sqrt{2}\,\psi'_1\,L_1 - \psi'_2\,L_2\right)\;,
\label{eq:Pviderivation}
\end{eqnarray}
which results in the K\"ahler correction
\begin{equation}
 \Delta \mathscr{K}~=~\kappa_{\rep{2'2'}}\,\,\left(\begin{array}{ccc}
2\,(\psi'_1)^2 & -\sqrt{2}\,(\psi'_1)^\dagger\,\psi'_2 & 0 \\
-\sqrt{2}\,\psi'_1\,(\psi'_2)^\dagger & (\psi'_1)^2+(\psi'_2)^2 & \sqrt{2}\,(\psi'_1)^\dagger\,\psi'_2 \\
0 & \sqrt{2}\,\psi'_1\,(\psi'_2)^\dagger & 2\,(\psi'_2)^2
\end{array}
\right) + \text{h.c.}\;.
\end{equation}
After inserting the VEV $\langle \psi'\rangle = \left(\psi'_0,\,
\psi'_0\right)^T$, we get 
$\Delta \mathscr{K} ~=~\kappa_{\rep{2'2'}}\,(\psi'_0/\Lambda)^2\,P + \text{h.c.}$ with
\begin{equation}
 P~=~ \left(\begin{array}{ccc}
2 & -\sqrt{2} & 0 \\
- \sqrt{2} & 2 & \sqrt{2} \\
0 & \sqrt{2} & 2
\end{array}
\right)\;.
\end{equation}

Summarizing, the $\mathrm{T}'$ model admits the same corrections as in the \A4
case, described in \Eqref{eq:pmatrix1} and \Eqref{eq:pmatrix2}, and in addition
corrections proportional to the $P$ matrices
\begin{subequations}
\label{eq:KaehlerBasisTprime}
\begin{equation}
 P_\mathsf{i}~=~\diag(0,2,1)\;,\quad
 P_\mathsf{ii}~=~\diag(1,0,2)
 \quad\text{and}\quad
 P_\mathsf{iii}~=~\diag(2,1,0)
\label{eq:Tpmatrix1}
\end{equation}
coming from contractions of $L$ with $\psi$. Since
$\langle\psi\rangle=(\psi_0,0)$, their contribution in the K\"ahler potential is
proportional to $(\psi_0)^2$. Furthermore, K\"ahler corrections proportional to
the three matrices
\begin{eqnarray}
 P_\mathsf{iv} & = & \begin{pmatrix}
 2 & \sqrt{2} & -\sqrt{2} \\ \sqrt{2} & 2 & 0\\ -\sqrt{2} & 0 & 2
 \end{pmatrix}\;,
 \quad
 P_\mathsf{v}~=~\begin{pmatrix}
 2 & 0& \sqrt{2} \\ 0& 2 & -\sqrt{2} \\ \sqrt{2} & -\sqrt{2} & 2
 \end{pmatrix} \quad \text{and} \nonumber \\
 P_\mathsf{vi} & = & \begin{pmatrix}
 2 & -\sqrt{2} & 0 \\ -\sqrt{2} & 2 & \sqrt{2}\\ 0 & \sqrt{2} & 2
 \end{pmatrix}\;,
\label{eq:Tpmatrix2}
\end{eqnarray}
\end{subequations}
are all due to contractions with $\psi'$. Since
$\langle\psi'\rangle=(\psi'_0,\psi'_0)$, their contribution is proportional to
$(\psi'_0)^2$.

\subsection{General $\boldsymbol{P}$ matrices}
\label{sec:GeneralP}

In the general case of different flavon VEVs, or possibly a different symmetry
group, one can imagine that not all K\"ahler correction matrices $P$ can be
expressed as linear combinations of the $P_i$ in
equations~\eqref{eq:KaehlerBasis} and \eqref{eq:KaehlerBasisTprime}.
Therefore, there are in general more possible $P$ matrices. However,
since the K\"ahler corrections are Hermitean, one can express a general
$P$ matrix in terms of nine Hermitean basis matrices,
\begin{subequations}\label{eq:Pmatrices}
\begin{eqnarray}
 P_1~=~\begin{pmatrix}
 1 & 0 & 0 \\ 0 & 0 & 0\\ 0 & 0 & 0
 \end{pmatrix}\;,\quad
& P_2~=~\begin{pmatrix}
 0 & 1 & 0 \\ 1 & 0 & 0\\ 0 & 0 & 0
 \end{pmatrix}\;, \quad
& P_3~=~\begin{pmatrix}
 0 & 0 & 0 \\ 0 & 1 & 0\\ 0 & 0 & 0
 \end{pmatrix}\;,
\\
 P_4~=~\begin{pmatrix}
 0 & 0 & 1 \\ 0 & 0 & 0\\ 1 & 0 & 0
 \end{pmatrix}\;, \quad
& P_5~=~\begin{pmatrix}
 0 & 0 & 0 \\ 0 & 0 & 1\\ 0 & 1 & 0
 \end{pmatrix}\;, \quad
& P_6~=~\begin{pmatrix}
 0 & 0 & 0 \\ 0 & 0 & 0\\ 0 & 0 & 1
 \end{pmatrix}\;, 
\\
 P_7~=~\begin{pmatrix}
 0 & -\I & 0 \\ \I & 0 & 0\\ 0 & 0 & 0
 \end{pmatrix}\;, \quad
& P_8~=~\begin{pmatrix}
 0 & 0 & -\I \\ 0 & 0 & 0\\ \I & 0 & 0
 \end{pmatrix}\;, \quad
& P_9~=~\begin{pmatrix}
 0 & 0 & 0 \\ 0 & 0 & -\I\\ 0 & \I & 0
 \end{pmatrix}\;.
\end{eqnarray}
\end{subequations}
Then, e.g., the $P$ matrices in equations~\eqref{eq:KaehlerBasis} of our
 $\A4$ example can be expressed as
\begin{equation}
 P_\mathrm{I}~=~P_1\;, 
 \quad P_\mathrm{II}~=~P_3\;, 
 \quad P_\mathrm{III}~=~P_6\;, \quad 
 P_\mathrm{IV}~=~ \sum_{i=1}^{6} P_i\;, \quad 
 P_\mathrm{V}~=~ \sum_{i=7}^{9} (-1)^i\, P_i\;,
\end{equation}
respectively.

\subsection{Analytic formulae for K\"ahler corrections}
\label{sec:formulae}

In this section, we give a detailed account of the derivation of analytical
formulae for the corrections to the mixing parameters coming from the K\"ahler
potential, building on earlier publications~\cite{Lindner:2005as,Antusch:2007vw}.

\subsubsection{The general idea}
\label{sec:GenIdea}

Let us start by specifying the goal of our derivation. We assume
that we are given a model that makes predictions for the leptonic mixing
parameters without taking into account any terms in the K\"ahler potential
beside the canonical ones. That is, the superpotential alone predicts the lepton
masses, mixing angles and complex phases, and the K\"ahler potential has the form
shown in \Eqref{eq:Kcanonical}. We emphasize that it is irrelevant for the
following computations how precisely the prediction for the parameters of the
lepton sector is achieved. In principle, we only need the charged lepton Yukawa
matrix and the Majorana neutrino mass matrix as input. Furthermore, for
computational simplicity, we assume that the model has been transformed to a
basis where the charged lepton Yukawa matrix is diagonal. Hence, a complete set
of input parameters is given by the three charged lepton masses, the three
neutrino masses and the nine mixing parameters,\footnote{This also includes the
three  ``unphysical'' phases that are usually absorbed in the charged lepton
fields. The reasons for this will be explained in detail below.} which we
assume to be in the standard parametrization (cf.\
appendix~\ref{app:parametrization}).

After having specified the input, we now consider the same model but amended
with correction terms in the K\"ahler potential. That is, we allow for an
arbitrary K\"ahler potential for the left--handed lepton doublets
$L=(L_1,L_2,L_3)$ and the right--handed lepton singlets $R=(R_1,R_2,R_3)$,
\begin{equation}
 K~\supset~L^\dagger\,\mathscr{K}_L\,L+R^\dagger\,\mathscr{K}_R\,R\;,
\end{equation}
with the Hermitean matrices
\begin{equation}\label{eq:KL}
 \mathscr{K}_{L/R}~=~\mathbbm{1}+\Delta \mathscr{K}_{L/R}\;.
\end{equation}
Because of the corrections $\Delta \mathscr{K}_{L/R}$, $L$ and $R$ are not
canonically normalized. Since $\mathscr{K}_{L/R}$ are Hermitean and
positive, they can be rewritten as 
\begin{equation} 
 \mathscr{K}_{L/R}~=~H_{L/R}^\dagger H_{L/R}~=~H_{L/R}^2\\
\end{equation}
with Hermitean $H_{L/R}=H_{L/R}^\dagger$, such that the
canonically normalized fields are
\begin{subequations}\label{eq:renormalize}
\begin{eqnarray}
  L' & =&H_L\,L\;,\\
  R' &=&H_R\,R\;.
\end{eqnarray}
\end{subequations}

Since we assume the K\"ahler corrections $\Delta \mathscr{K}_{L/R}$ to
arise from terms that are suppressed with respect to the canonical terms by
powers of the flavon VEVs over the fundamental scale, we specialize to
infinitesimal $\Delta\mathscr{K}_{L/R}$,
\begin{subequations}
\begin{eqnarray}\label{eq:Pmatrix}
 \Delta\mathscr{K}_L & = & -2\,x_1\,P_L\;,\\
 \Delta\mathscr{K}_R & = & -2\,x_2\,P_R\;.
\end{eqnarray}
\end{subequations}
Here $P_L$ and $P_R$ are Hermitean, $x_1$ and $x_2$ are infinitesimal, and the
factors $-2$ turn out to be convenient. The goal of the following discussion is
to find analytic formulae for the dependence of the mixing parameters on $x_1$
and $x_2$ for generic $P_{L/R}$.

Before we go on with the discussion, a comment is in order. We assume that the
lepton basis is chosen such that the charged lepton Yukawa matrix is diagonal
with positive real entries. However, this does not fix the basis completely
since one can still perform a phase redefinition of the left--handed and
right--handed lepton fields without changing the Yukawa matrix as long as the
phase change is the same for both sectors. This freedom is conventionally used
to absorb the three so--called unphysical phases $\delta_e$, $\delta_\mu$ and
$\delta_\tau$ of $U_\mathrm{PMNS}$ into the fields. However, this is only
possible for a diagonal K\"ahler potential. In the non--diagonal case discussed
here, this redefinition also changes $P_{L/R}$ according to
\begin{equation}
\label{eq:Pprime}
  P_{L/R} \rightarrow \widehat{P}_{L/R}~=~U_\mathrm{ph}^\dagger\, P_{L/R} \,U_\mathrm{ph}\;,
\end{equation}
where $U_\mathrm{ph}=\diag{(e^{\I \delta_e},e^{\I \delta_\mu},e^{\I
\delta_\tau})}$ is a diagonal matrix containing the unphysical phases. That is,
after setting the unphysical phases to zero in the mixing matrix, the
transformed $\widehat{P}_{L/R}$ depends on them. This shows explicitly that the
changes in the mixing parameters depend on the values of $\delta_e$,
$\delta_\mu$ and $\delta_\tau$.\footnote{This does not imply that these phases
are physical in this case. After the K\"ahler potential has been diagonalized,
one can choose a field basis in which $\delta_{e, \mu, \tau}$ are zero.} Our results are derived
for the case in which the three phases in $U_\mathrm{PMNS}$ are zero at
$x_1=x_2=0$. If this is not the case, one has to apply the resulting formulae
not to the original $P_{L/R}$ but to $\widehat{P}_{L/R}$ as defined in
\Eqref{eq:Pprime}.

For the following discussion, it turns out to be useful to introduce for any
unitary matrix $U$ a corresponding anti--Hermitean matrix 
\begin{equation}
 T~:=~U^\dagger U'\;,
\end{equation}
where the prime denotes the derivative with respect to $x_1$ or $x_2$ (it will be clear from the context which one is meant), such that
\begin{equation}
 U' ~=~U\,T\,.
\end{equation}
Hence, for the leptonic mixing matrix $U_\mathrm{PMNS}$, we define
\begin{equation}\label{eq:TPMNSdef}
  T_\mathrm{PMNS}~:=~U_\mathrm{PMNS}^\dagger\, U_\mathrm{PMNS}'\;.
\end{equation}
Since $T_\mathrm{PMNS}$ is anti--Hermitean, it has nine independent real parameters
\begin{equation}
 u~:=~\left\{\re T_{12},\re T_{13},\re T_{23},
 \im T_{11},\im T_{12},\im T_{13},\im T_{22},\im T_{23},\im T_{33}
 \right\}\;. 
\end{equation}
On the other hand, one can use the definition of $T_\mathrm{PMNS}$, i.e.\ \Eqref{eq:TPMNSdef}, in order
to express its entries in terms of the mixing parameters and their derivatives.
Clearly, $T_\mathrm{PMNS}$ is linear in the derivatives of the mixing parameters.
Therefore, there is a linear map $A$ 
from the derivatives of the mixing parameters to the elements of $T_\mathrm{PMNS}$, 
which only depends on the mixing parameters but not on their derivatives. 
Defining
\begin{equation}
 \xi~:=~
 \{\theta_{12}', \theta_{13}', \theta_{23}', 
 \delta', \delta_{e}', \delta_{\mu}', \delta_{\tau}', 
 \varphi_1', \varphi_2'\}\;,
\end{equation}
one can write this relation in matrix form,
\begin{equation}
 A\,\xi~=~u\;.
\end{equation}
The first steps of our computation are to compute $u$ in terms of the mixing parameters, then to read off $A$ from this expression and finally to invert $A$.
This way one obtains linear differential equations for the mixing parameters. The remaining task is to find $T_\mathrm{PMNS}$ for arbitrary given K\"ahler corrections $P_{L/R}$.

One can split this task into several parts. First, one can make use of the
definition of $U_\mathrm{PMNS}:=U_e^\dagger U_\nu$ in order to split the
derivative $U'_\mathrm{PMNS}$ into two parts,
\begin{equation}
  U'_\mathrm{PMNS}~=~(U_e')^\dagger U_\nu + U_e^\dagger U_\nu'\,.
\end{equation}
Multiplying this with $U_\mathrm{PMNS}^\dagger$ from the left and inserting
twice an identity matrix yields
\begin{equation}\label{eq:TPMNSexpand}
  T_\mathrm{PMNS}~=~U_\mathrm{PMNS}^\dagger \, (U_e')^\dagger \, (U_e U_e^\dagger) \, U_\nu + U_\mathrm{PMNS}^\dagger \, U_e^\dagger \, (U_\nu U_\nu^\dagger) \, U_\nu'\;.
\end{equation}
This can be further simplified by the introduction of the matrices
$T_e=U_e^\dagger U_e'$ and $T_\nu=U_\nu^\dagger U_\nu'$ and because of the
anti--Hermiticity of $T_e$ one finally arrives at
\begin{equation}
  T_\mathrm{PMNS}~=~T_\nu - U_\mathrm{PMNS}^\dagger \, T_e \, U_\mathrm{PMNS}\;.
\end{equation}
In the following, we compute the different contributions to $T_\nu$ and $T_e$ at
$x_1=x_2=0$. All quantities are from now on evaluated at this point if not
indicated otherwise by displaying the arguments $(x_1)$ or $(x_2)$ explicitly.

\subsubsection{Corrections due to \boldmath $U_\nu$ \unboldmath}

Let us first discuss the case of $T_\nu$. Going to canonically normalized fields
by the transformation shown in \Eqref{eq:renormalize} leads to the
change of the neutrino mass matrix 
\begin{eqnarray}
 \mathscr{W}_\nu
 & = & 
 \frac{1}{2}L^T\,m_\nu^0\,L
 \nonumber\\
 & \simeq & 
 \frac{1}{2}\left[\left(\mathbbm{1}+x_1\,P_L\right)\,L'\right]^T\,m_\nu^0\,
 	\left[\left(\mathbbm{1}+x_1\,P_L\right)\,L'\right]
 \nonumber\\
 & \simeq & 
 \frac{1}{2}L^{\prime\,T}\,m_\nu^0\,L'
 + \frac{1}{2}
 x_1\,L^{\prime\,T}\,\left(P_L^T\,m_\nu^0+m_\nu^0\,P_L\right)\,L'
\end{eqnarray}
in linear order in $x_1$.
That is, the change of $m_\nu$ is governed by a differential equation which has
the same form as the renormalization group equation (RGE) for the neutrino mass
operator (cf.\ equation~(B.5) of \cite{Antusch:2003kp}),
\begin{equation}
 \frac{\D}{\D x_1}m_\nu(x_1)~=~P^T\,m_\nu^0+m_\nu^0\,P\;, \qquad m_\nu(0)=m_\nu^0\;.
\end{equation}
Therefore, we can repeat the steps in \cite{Antusch:2003kp} 
that have led to the analytic solutions to 
the RGEs for the mixing parameters.

The $x_1$--dependent unitary diagonalization matrix $U_\nu(x_1)$ of the neutrino
mass matrix $m_\nu(x_1)$ is defined by the equation,
\begin{equation}
\label{eq:Diagonalisation}
 U_\nu(x_1)^T\,m_\nu(x_1)\,U_\nu(x_1)~=~D_\nu(x_1)~=~\diag\left(m_1(x_1),m_2(x_1),m_3(x_1)\right)\;,
\end{equation}
where the mass eigenvalues $m_i(x_1)$ are positive real numbers.

Taking the derivative of \Eqref{eq:Diagonalisation}  with respect to $x_1$, 
which is denoted by a prime in the
following, and evaluating the result at $x_1=0$,
we obtain
\begin{eqnarray}
 \frac{\D}{\D x_1}\left(U_\nu^*(x_1)\,D_\nu(x_1)\,U_\nu^\dagger(x_1)\right)|_{x_1=0}
 & = & 
 (U_\nu')^*\,D_\nu\,U_\nu^\dagger
 +
 U_\nu^*\,D_\nu\,(U_\nu')^\dagger
 +
 U_\nu^*\,D_\nu'\,U_\nu^\dagger
 \nonumber\\
 & = & P_L^T\,U_\nu^*\,D_\nu\,U_\nu^\dagger+U_\nu^*\,D_\nu\,U_\nu^\dagger\,P_L\;.
\end{eqnarray}
All quantities on the right--hand side are evaluated at $x_1=0$. Multiplying
this equation by $U_\nu^T$ from the left and by $U_\nu$ from the right yields
\begin{equation}
 U_\nu^T\,(U_\nu')^*\,D_\nu
 +
 D_\nu\,(U_\nu')^\dagger\,U_\nu
 +
 D_\nu'
 ~=~
 \widetilde{P}_L^T\,D_\nu+D_\nu\,\widetilde{P}_L
\end{equation}
with 
\begin{equation}
 \widetilde{P}_L~=~U_\nu^\dagger\,P_L\,U_\nu~=~U_\mathrm{PMNS}^\dagger\,P_L\,U_\mathrm{PMNS}\;,
\end{equation} 
where we used the fact that $U_e(0,0)=\mathbbm{1}$.
With the previously defined anti--Hermitean matrix $T_\nu=U_\nu^\dagger U_\nu'$
one can rewrite this equation as
\begin{equation}
 D_\nu'~=~\widetilde{P}_L^T\,D_\nu+D_\nu\,\widetilde{P}_L
 -T_\nu^*\,D_\nu+D_\nu\,T_\nu\;.
\end{equation}
Since the left--hand side of this equation is diagonal and real, the right--hand
side has to have these properties as well and one obtains
\begin{equation}
 m_i'~=~2\,(\widetilde{P}_L)_{ii}\,m_i+\bigl((T_\nu)_{ii}-(T_\nu^*)_{ii}\bigr)\, m_i\;.
\end{equation}
The first term is real since $P_L$ (and thus 
$\widetilde{P}_L$) is Hermitean, whereas the second term is purely imaginary and has to
vanish,
\begin{equation}\label{eq:Tii}
 \im (T_\nu)_{ii}~=~0\;.
\end{equation}
By comparing the off--diagonal terms one gets
\begin{equation}
 m_i\,(T_\nu)_{ij}-(T_\nu)_{ij}^*\,m_j
 ~=~
 -(\widetilde{P}_L^T)_{ij}\,m_j-m_i\,(\widetilde{P}_L)_{ij}
\end{equation}
such that
\begin{subequations}
\begin{eqnarray}
 \re (T_\nu)_{ij}
 & = & 
 -\frac{m_j\,\re (\widetilde{P}_L)_{ji}+m_i\,\re (\widetilde{P}_L)_{ij}}{m_i-m_j}~=~-\frac{m_i+m_j}{m_i-m_j}\,\re (\widetilde{P}_L)_{ij}
 \;,\\
 \im (T_\nu)_{ij}
 & = & 
 -\frac{m_j\,\im (\widetilde{P}_L)_{ji}+m_i\,\im (\widetilde{P}_L)_{ij}}{m_i+m_j}~=~-\frac{m_i-m_j}{m_i+m_j}\,\im (\widetilde{P}_L)_{ij}
 \;.
\end{eqnarray}
\end{subequations}
This, together with \eqref{eq:Tii}, specifies all entries of $T_\nu$.

\subsubsection{Corrections due to \boldmath $U_e$ \unboldmath}

We turn now to corrections coming from the charged lepton sector.
The Yukawa coupling term changes due to the redefinition of the fields as given
in \Eqref{eq:renormalize},
\begin{eqnarray}
  \notag \mathscr{W}_e & = & -R^T \, Y_e^0 \, L + \mathrm{h.c.}\\
  \notag & = & -R^{\prime\,T}\, \left((H_R)^{-1}\right)^T \, Y_e^0 \, (H_L)^{-1} \, L' + \mathrm{h.c.}\\
  \notag & \simeq & -R^{\prime\,T}\, (\mathbbm{1} + x_2 \, (P_R)^T)\, Y_e^0 \, (\mathbbm{1} + x_1 \, P_L) \, L' + \mathrm{h.c.}\\
  & \simeq & -R^{\prime\,T}\, (Y_e^0 + x_2 \, (P_R)^T\, Y_e^0 + x_1 \, Y_e^0 \, P_L) \, L' + \mathrm{h.c.}\;.
\end{eqnarray}
The Yukawa matrix can be diagonalized by a bi--unitary transformation. The
matrix $U_e$ acting on the left--handed charged leptons is determined to first
order in $x_1$ and $x_2$ by
\begin{multline}\label{eq:Ue}
  U_e^\dagger(x_1,x_2)\, \left[(Y_e^0)^2 + 2\,x_2 \, Y_e^0 \, (P_R)^T\, Y_e^0 + x_1 \, (Y_e^0)^2 \, P_L + x_1 \, P_L \, (Y_e^0)^2 \right] U_e(x_1,x_2)\\
  =~D_e^2(x_1,x_2)~=~\diag{(y_1^2(x_1,x_2),y_2^2(x_1,x_2),y_3^2(x_1,x_2))}\;,
\end{multline}
where $D_e^2(x_1,x_2)$ is the diagonal matrix of the squared lepton Yukawa
couplings and $D_e^2(0,0)=(Y_e^0)^2$.

We first focus on the $x_1$--dependence. Hence, we take the derivative of
\Eqref{eq:Ue} with respect to $x_1$ and evaluate the resulting expression at
$x_1=0$, $x_2=0$ (remember that $U_e(0,0)=\mathbbm{1}$),
\begin{eqnarray}
 \frac{\D}{\D x_1}\left(U_e(x_1)\,D_e^2(x_1)\,U_e^\dagger(x_1)\right)|_{x_1=0}
 & = & 
 U_e'\,(Y_e^0)^2
 +
 (Y_e^0)^2\,(U_e')^\dagger
 +
 (D_e^2)'
 \nonumber\\
 & = & P_L\,(Y_e^0)^2+(Y_e^0)^2\,P_L\;.
\end{eqnarray}
Using the fact that $T_e^{x_1}=U_e'$ (where the superscript $x_1$ on $T_e^{x_1}$
means that it is the part of $T_e$ which corresponds to changes of $x_1$) is
anti--Hermitean, one can rearrange this equation into
\begin{equation}\label{eq:Tex}
 (D_e^2)'~=~
  P_L\,(Y_e^0)^2
  + (Y_e^0)^2\,P_L 
 - T_e^{x_1}\,(Y_e^0)^2
 + (Y_e^0)^2\,T_e^{x_1}\;.
\end{equation}
For the terms on the diagonal this reads
\begin{equation}
 (y_i^2)'~=~2\,(P_L)_{ii}\,y_i^2\;.
\end{equation}
There is one important difference from the case of $U_\nu$. The diagonal terms
$(T_e^{x_1})_{ii}$ cancel exactly; hence, they cannot be determined from this
equation. However, they only contribute to the changes of the unphysical phases
which can, after diagonalizing the K\"ahler potential, be transformed
away.\footnote{We have verified this analytically by keeping the diagonal
entries of $T_e$ arbitrary.} Therefore, changes of these phases are of no
interest to us. Contrary to the diagonal terms, the off--diagonal terms of
$T_e^{x_1}$ can be derived from \Eqref{eq:Tex},
\begin{equation}
  (T_e^{x_1})_{ij}= (P_L)_{ij} \, \frac{y_i^2+y_j^2}{y_j^2 - y_i^2}\;.
\end{equation}
This fixes the first part of $T_e^{x_1}$ up to the imaginary parts of the
entries on the diagonal.

Moreover, there is the part of $T_e$ which comes from changes in $x_2$, denoted
by $T_e^{x_2}$. Besides the fact that the derivatives are now taken with respect
to $x_2$, the only difference compared to \Eqref{eq:Tex} is that $P_L\,(Y_e^0)^2
+ (Y_e^0)^2\,P_L $ is replaced by $2\, Y_e^0 \, (P_R)^T\, Y_e^0$, i.e.\ 
\begin{equation}\label{eq:Tey}
 (D_e^2)'~=~ 2\, Y_e^0 \, (P_R)^T\, Y_e^0 - T_e^{x_2}\,(Y_e^0)^2 +
 (Y_e^0)^2\,T_e^{x_2}\;.
\end{equation}
From this equation one can derive the change of the masses,
\begin{equation}
 (y_i^2)'~=~2\,(P_R)_{ii}\,y_i^2\;,
\end{equation}
which again leaves the terms on the diagonal of $T_e^{x_2}$ undetermined,
and the off--diagonal terms of $T_e^{x_2}$,
\begin{equation}
  (T_e^{x_2})_{ij}= 2\,(P_R)_{ji} \, \frac{y_i\, y_j }{y_j^2 - y_i^2}\;.
\end{equation}
This determines $T_e^{x_2}$ up to the entries on the diagonal.

Combining all three contributions yields
\begin{subequations}
\begin{eqnarray}
  T_\mathrm{PMNS}^{x_1} & = & -U_\mathrm{PMNS}^\dagger \, T_e^{x_1} \, U_\mathrm{PMNS} + T_\nu\;,\\
  T_\mathrm{PMNS}^{x_2} & = & -U_\mathrm{PMNS}^\dagger \, T_e^{x_2} \, U_\mathrm{PMNS}\;.
\end{eqnarray}
\end{subequations}
From this, one can derive the derivatives of the mixing parameters at $x_1=0$,
$x_2=0$ from which the change of the parameters can be computed to first order
by simple multiplication with $x_1$ and $x_2$, respectively.

\subsection{\texttt{Mathematica} package}
\label{sec:Mathematica}

The formulae that we obtained by following the procedure outlined above
are made available online as a \texttt{Mathematica} package. It can be found on
the web--page
\begin{center}
 \url{http://einrichtungen.ph.tum.de/T30e/codes/KaehlerCorrections/}.
\end{center}
The package contains the full analytic formulae in such a way that all
input parameters can be set by the user. Without specifying any initial values,
the formulae are very lengthy, and therefore, we will in the following only use
them after setting most of the input parameters (cf.\
appendix~\ref{app:ExForm}).
   
Furthermore, the package provides some functions to simplify the usage of the
formulae. In particular, the function \texttt{kaehlerCorr} can be used to output
the K\"ahler corrections for a K\"ahler potential of the form
\begin{equation}
  K~=~
  \left(L^f\right)^\dagger \, \left(\delta_{fg} + x_L \, (P_L)_{fg}\right) \,L^g 
  + \left(R^f\right)^\dagger \, \left(\delta_{fg} + x_R \, (P_R)_{fg}\right) \,R^g
  \;,
\end{equation}
with $P_L$, $P_R$ fixed and given initial values for masses and mixing
parameters. We emphasize again that the formulae may only be applied in a
basis where the charged lepton Yukawa matrices are diagonal.

Some care is to be exercised for the case of a zero initial
mixing angle as it occurs in tri--bi--maximal mixing because this renders the
initial value of $\delta$ undefined. One can infer its correct value by the
requirement that the change of $\delta$ is analytical in the angle that has zero
initial value. This is done automatically by the package. However, there is also
the possibility to override this behaviour in case the automatic determination
fails. For more information, we refer the reader to the documentation which is part of the
download.

\section{Implications}
\label{sec:implications}

After presenting the derivation of analytic formulae for corrections from the
K\"ahler potential, we now apply these formulae to explicit examples from the
literature, focussing on the models based on \A4 \cite{Altarelli:2005yx}
and on $\mathrm{T}'$ \cite{Chen:2009gf} introduced in
\Secref{sec:HolomorphicTerms}. We also discuss the implications of the K\"ahler
corrections for the VEV alignment and the threat of flavor changing neutral
currents. First, however, we present the changes induced by the K\"ahler
corrections starting from tri--bi--maximal and bi--maximal mixing for general
$P$ matrices from \Eqref{eq:Pmatrices}.

\subsection{Tables for general \boldmath $P$ matrices \unboldmath }

We now provide tables that summarize the K\"ahler corrections starting
from tri--bi--maximal mixing and bi--maximal mixing, respectively. For the
charged lepton masses and the mass--squared differences of the neutrinos, 
the current PDG values \cite{Beringer:1900zz} are used and normal hierarchy is
assumed. The absolute mass scale of the neutrinos is set by $m_1=0.01\ev$. The
form of the K\"ahler potential under consideration is
\begin{equation}
  K~=~L^\dagger\, (1+x_L\, P_L)\, L + R^\dagger\, R\;,
\end{equation}
i.e.\ only the left--handed sector is modified. In the tables, $P_L$ is replaced
by one of the nine basis matrices $P_i$, see \Eqref{eq:Pmatrices}, in each
column. The value of the small parameter is $x_L=0.01$. The results are
summarized in \Tabref{tab:tribi} for tri--bi--maximal mixing, i.e.\
\begin{eqnarray}
  \theta_{12} & = & \arcsin{\frac{1}{\sqrt{3}}}\;, \quad 
  \theta_{13}~=~0,\quad 
  \theta_{23}~=~\frac{\pi}{4}\;,\quad 
  \delta~=~\text{undefined}\;,  \nonumber \\ 
  \delta_e & = & \pi\;,\quad 
  \delta_\mu~=~\pi\;,\quad  
  \delta_\tau~=~0\;,\quad 
  \varphi_1~=~\varphi_2~=~2\pi\;,
\end{eqnarray}
where the phases are determined from \Eqref{eq:TBMmatrix}, and in
\Tabref{tab:bi} for bi--maximal mixing, i.e.\
\begin{eqnarray}
 \theta_{12} & =& \frac{\pi}{4}\;,\quad  
 \theta_{13}~=~0\;,\quad 
 \theta_{23}~=~\frac{\pi}{4}\;,\quad 
 \delta~=~\text{undefined}\;,  \nonumber \\ 
 \delta_e & = & \pi\;,\quad  
 \delta_\mu~=~\pi\;,\quad \delta_\tau~=~0\;,\quad 
 \varphi_1~=~\varphi_2~=~2\pi\;,
\end{eqnarray}
where, for simplicity, the same phases have been chosen. In
appendix~\ref{app:ExForm} we also present the analytic formulae for
tri--bi--maximal mixing but without setting absolute neutrino masses and without
specifying $x_L$.

We should emphasize that the results shown in these tables depend on all the
mixing parameters before taking into account the K\"ahler corrections. That is,
in particular, they also depend on the two Majorana phases and the three
unphysical phases. Furthermore, at the starting points of tri--bi--maximal and
bi--maximal mixing, the phase $\delta$ is not properly defined due to
$\theta_{13}=0$. For each $P_i$, it is determined from the formulae by demanding
that the change of $\delta$ is analytical at $\theta_{13}=0$.

\begin{table}[t!]
\centering
\begin{tabular}{rccccccccc} \hline
  & $P_1$ & $P_2$ & $P_3$ & $P_4$ & $P_5$ & $P_6$ & $P_7$ & $P_8$ & $P_9$ \\ \hline
  $\Delta \theta_{12}$ $[^\circ]$: & -0.96 & -0.28 & 0.48 & -0.28 & 0.96 & 0.48 & 0 & 0 & 0\\
  $\Delta \theta_{13}$ $[^\circ]$: & 0 & -0.12 & -0.015 & 0.12 & 0 & 0.015 & -0.073 & 0.073 & 0.012\\
  $\Delta \theta_{23}$ $[^\circ]$: & 0 & -0.021 & -0.24 & 0.021 & -0.29 & 0.24 & 0 & 0 & 0 \\ \hline
\end{tabular}
\caption{Changes of the mixing angles under K\"ahler corrections of the form $\Delta K=x \, L^\dagger \, P_i \, L$ for $x=0.01$ (cf.\ \Eqref{eq:Pmatrices}) starting from tri--bi--maximal mixing.}
\label{tab:tribi}
\end{table}

\begin{table}[t!]
\centering
\begin{tabular}{rccccccccc} \hline
  & $P_1$ & $P_2$ & $P_3$ & $P_4$ & $P_5$ & $P_6$ & $P_7$ & $P_8$ & $P_9$ \\ \hline
  $\Delta \theta_{12}$ $[^\circ]$: & -1.0 & 0.20 & 0.51 & 0.20 & 1.0 & 0.51 & 0 & 0 & 0\\
  $\Delta \theta_{13}$ $[^\circ]$: & 0 & -0.12 & -0.016 & 0.12 & 0 & 0.016 & -0.076 & 0.076 & 0.012\\
  $\Delta \theta_{23}$ $[^\circ]$: & 0 & -0.023 & -0.23 & 0.023 & -0.29 & 0.23 & 0 & 0 & 0 \\ \hline
\end{tabular}
\caption{Changes of the mixing angles under K\"ahler corrections of the form $\Delta K=x \, L^\dagger \, P_i \, L$ for $x=0.01$ (cf.\ \Eqref{eq:Pmatrices}) starting from bi--maximal mixing.}
\label{tab:bi}
\end{table}

\subsection{Corrections in the \boldmath \A4 \unboldmath model}
\label{sec:corrA4}

\begin{figure}
  \centering
  \includegraphics{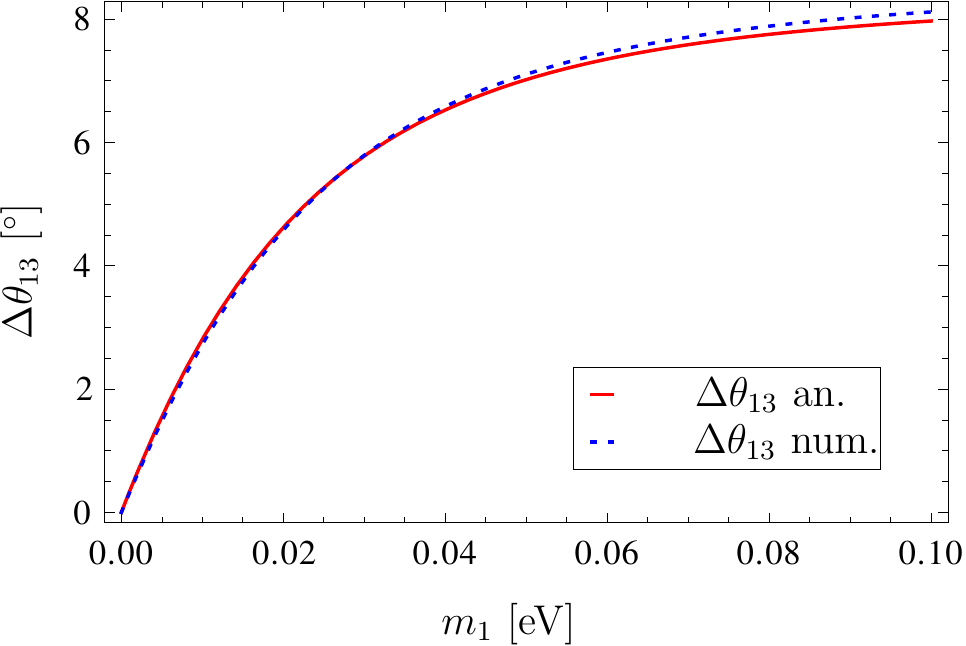}
  \caption{Change of $\theta_{13}$ in the \A4 model due to the K\"ahler correction from the matrix $P_\mathrm{V}$, setting $\kappa_\mathrm{V}\,v^2/\Lambda^2=(0.2)^2$. The continuous line shows the result of \Eqref{eq:theta13An}, while the dashed line represents the result of a numerical computation.}
  \label{fig:theta13}
\end{figure}

We start with a discussion of the K\"ahler corrections in the \A4
model. As shown in section \ref{sec:quadratic}, there are five independent
quadratic corrections which cannot be forbidden by a symmetry. The matrix
$P_\mathrm{V}$, for example, comes from the higher--order term  $(L \otimes
\AfourFlavonA)^{\dagger}_{\rep{3}_s}(L \otimes \AfourFlavonA)_{\rep{3}_a}$,
as shown in \Eqref{eq:PVderivation}. 
If we plug $P_\mathrm{V}$ into our
derivation of the analytic formulae, we obtain for the change of $\theta_{13}$
from its tri--bi--maximal starting value the formula~\cite{Chen:2012ha} (cf.\ appendix~\ref{app:ExForm})
\begin{eqnarray}
\label{eq:theta13An}
      \notag \Delta \theta_{13} & = &\kappa_\mathrm{V} \cdot \frac{v^2}{\Lambda^2} \cdot 3 \sqrt{3} \cdot \frac{1}{\sqrt{2}} \left( \frac{2 m_1}{m_1+m_3} + \frac{m_e^2}{m_\mu^2-m_e^2} + \frac{m_e^2}{m_\tau^2-m_e^2}\right)\\
      & \simeq & \kappa_\mathrm{V} \cdot \frac{v^2}{\Lambda^2} \cdot 3\sqrt{6} ~ \frac{m_1}{m_1+m_3}\;,
\end{eqnarray}
assuming in the last line that the small contribution of the charged leptons can
be neglected. Using the PDG \cite{Beringer:1900zz} values for the mass--squared
differences, we can plot the change in $\theta_{13}$ against the neutrino mass
$m_1$ as shown in \Figref{fig:theta13}, where we set the ratio of VEV to the
cut--off scale to be of the order of the Cabibbo angle,
i.e.\ $v/\Lambda = 0.2$, and the coefficient $\kappa_\mathrm{V}=1$.
Unlike $\theta_{13}$, which approaches $\Delta\theta_{13}\approx 8.42^\circ$ for
$m_1 \rightarrow \infty$, the other angles experience only minor changes under
the correction $P_\mathrm{V}$. We see that with this correction we get close to
realistic values for $\theta_{13}$ while the other angles stay almost the
same.\footnote{To be precise, the \A4 model presented in
\Secref{sec:HolomorphicTerms} does not allow for a variation of $m_1$ while
keeping the mass--squared differences fixed. This is, however, possible in
extended models leading to tri--bi--maximal mixing, see e.g.\
\cite{Grimus:2009pg}.}

However, we also observe opposite effects, i.e.\ corrections that drive the
predictions of the angles away from their best fit values. For instance, the
corrections due to $P_\mathrm{IV}$, which are independent of the neutrino
masses, leave $\theta_{13}$ unchanged and change $\theta_{12}$ by $+3.2^\circ$
and $\theta_{23}$ by $-2.3^\circ$, given that we set
$\kappa_\mathrm{IV}\,v^2/\Lambda^2=(0.2)^2$. We have cross--checked these
analytical results by a numerical computation.

\subsection{Corrections in the \boldmath $\mathrm{T}'$ \unboldmath model}

In \Secref{sec:tprimeQu} we already described the  quadratic correction terms
for a $\mathrm{T}'$ model. As discussed there, due to its flavon structure the
model includes all of the correction terms of the \A4 model, so some of the
discussion from \Secref{sec:corrA4} still applies. However, the considered
$\mathrm{T}'$ model~\cite{Chen:2009gf} does not predict exact tri--bi--maximal 
mixing so we have to consider different initial values in our analytic formulae.
Moreover, a crucial assumption for the applicability of our formulae is that the
model is in a basis where the charged lepton Yukawa matrix is diagonal, as
stated in \Secref{sec:GenIdea}, which is also not the case in the considered
$\mathrm{T}'$ model. Therefore, we first have to perform a basis transformation
such that the charged lepton Yukawa matrix becomes diagonal. Since this is
simply a basis transformation, the mixing matrix and, hence, the mixing angles
are not affected. Nevertheless, the form of our correction matrices $P$ changes
which we demonstrate with the help of an example. The higher--order K\"ahler
potential term $(L \otimes \AfourFlavonA)^{\dagger}_{\rep{3}_s}(L \otimes
\AfourFlavonA)_{\rep{3}_a}$, after VEV insertion, leads to the term
\begin{equation}
 \Delta K~\supset~L^\dagger\,P_\mathrm{V}\,L+\text{h.c.}\;.
\end{equation}
However, this is in a basis where the charged lepton Yukawa matrix is
non--diagonal. The necessary basis transformation that diagonalizes it
redefines the left--handed charged leptons $L$ by some matrix $V$. This leads to
a modified $P$ matrix
\begin{equation}
L^\dagger\,P_\mathrm{V}\,L~\rightarrow~
(V\,L)^\dagger\,P_\mathrm{V}\,V\,L\,~=~ L^\dagger\,\widetilde{P}_\mathrm{V}\,L\;,
\end{equation}
where we defined $\widetilde{P}_\mathrm{V}:=V^\dagger\,P_\mathrm{V}\,V$.

\begin{figure}[t!]
  \centering
  \includegraphics{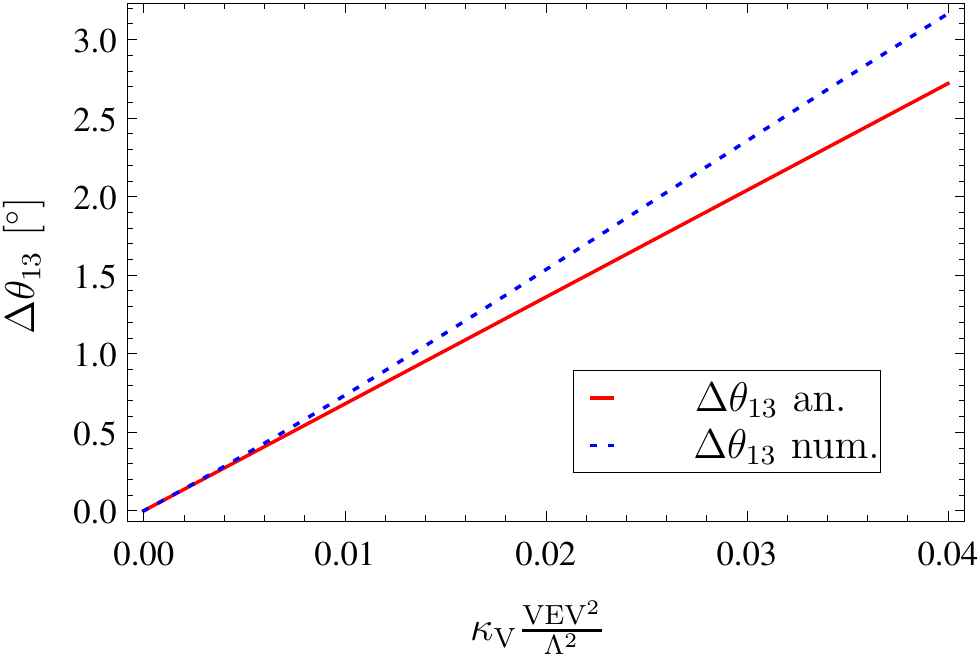}
  \caption{Change of $\theta_{13}$ in the $\mathrm{T}'$ model due to the K\"ahler correction from the (modified, see text) matrix $P_\mathrm{V}$ up to $\kappa_\mathrm{V}\,\mathrm{VEV}^2/\Lambda^2=(0.2)^2$. The continuous line shows the result of the analytical formulae while the dashed line represents the result of a numerical computation.}
  \label{fig:PVtheta13Tprime}
\end{figure}

In this basis we can now use our analytic formulae on the matrix
$\widetilde{P}_\mathrm{V}$, using the initial values for the mixing angles  
predicted by the original model as shown in~\Eqref{eq:TprimeStart}, 
$\theta_{12}\approx33^\circ, \theta_{23}=45^\circ$ and
$\theta_{13}\approx3^\circ$. Furthermore, we have to consider that the
model also predicts absolute neutrino masses, e.g.\ $m_1=0.0156\,\Ev$.
Therefore, we cannot plot the change in mixing angles as a function of the
neutrino masses, but rather against the size of the small expansion parameter
$x=\mathrm{VEV}^2/\Lambda^2$ times a coefficient $\kappa_\mathrm{V}$ from the
K\"ahler potential. For $\theta_{13}$ this is shown
in~\Figref{fig:PVtheta13Tprime}.

In this plot we see that $\theta_{13}$ can be increased by $\Delta
\theta_{13}\approx3^\circ$ for $\mathrm{T}'$, raising the value of $\theta_{13}$
up to about $6^\circ$ in the assumed parameter range. We also should comment
that in this model two flavon triplets have the same VEV structure, as one can
see in~\Tabref{tab:tprime}. According to \Eqref{eq:tprimeVEV}, both flavons
$\phi$ and $\xi$ have VEVs proportional to $\left(1, 1, 1\right)^T$ and,
therefore, can lead to the correction $P_\mathrm{V}$. In the best case, both
corrections would add up and boost the maximal change to
$\Delta\,\theta_{13}\approx6^\circ$. This would yield
$\theta_{13}\approx9^\circ$ as a result, which is of the order of the
experimentally measured value. However, this only applies to the very special
situation in which the contributions from both flavons $\phi$ and $\xi$ add and
contributions different from $P_\mathrm{V}$ should not spoil the result.
Moreover, the VEVs in the model are generally such that
$\mathrm{VEV}^2/\Lambda^2\sim \mathcal{O}(1/1000)$, in which case the K\"ahler
corrections become negligible.

\begin{figure}[t!]
  \centering
  \includegraphics{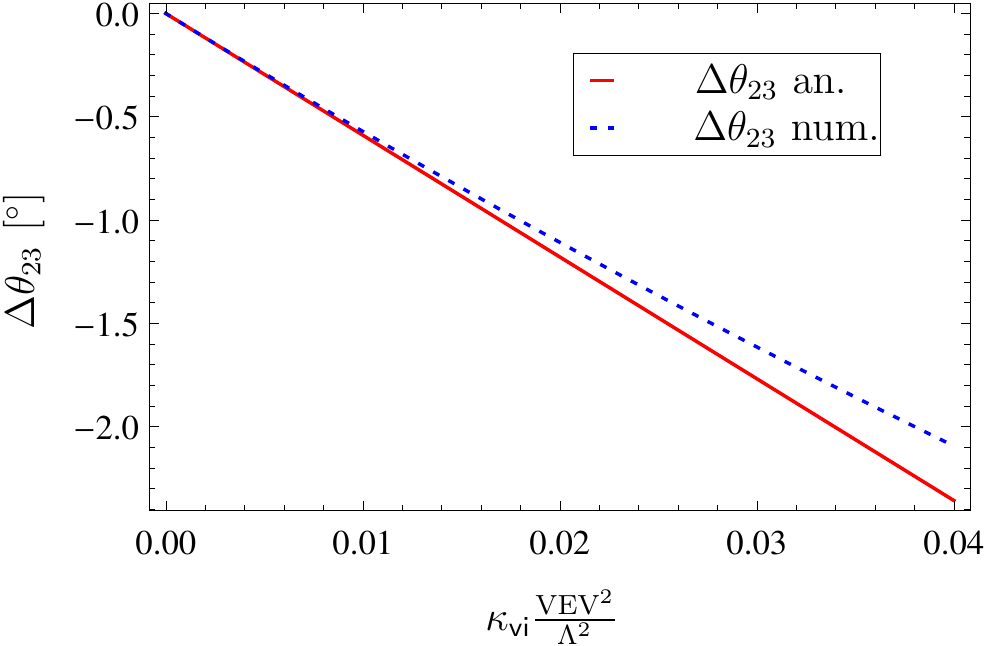}
  \caption{Change of $\theta_{23}$ in the $\mathrm{T}'$ model due to the K\"ahler correction from the (modified, see text) matrix $P_\mathsf{vi}$ up to $\kappa_\mathsf{vi}\,\mathrm{VEV}^2/\Lambda^2=(0.2)^2$. The continuous line shows the result of the analytical formulae while the dashed line represents the result of a numerical computation.}
  \label{fig:Pdoublet6theta23Tprime}
\end{figure}

In addition to the corrections which are also present in \A4, the $\mathrm{T}'$
model has, as we showed in \Secref{sec:tprimeQu}, six independent corrections
due to the flavon doublets $\psi$ and $\psi'$. Let us, for example, consider the
correction due to the matrix $P_\mathsf{vi}$ in \Eqref{eq:Tpmatrix2}, which
comes from the K\"ahler potential term $(L \otimes \psi')^{\dagger}_{\rep{2'}}(L
\otimes \psi')_{\rep{2'}}$ as can be seen in \Eqref{eq:Pviderivation}. Before we
can calculate the associated correction, we again have to perform a basis
transformation which brings the charged lepton Yukawa matrix into diagonal form,
therefore, also transforming $P_\mathsf{vi} \rightarrow
\widetilde{P}_\mathsf{vi}$. Using this matrix, the initial values
from~\Eqref{eq:TprimeStart} and the computed neutrino masses, we can again plot
the changes of the mixing angles against $x=\mathrm{VEV}^2/\Lambda^2$ times a
coefficient $\kappa_\mathsf{vi}$. As an example, we show the result for
$\theta_{23}$ in~\Figref{fig:Pdoublet6theta23Tprime}.

We hence see that the K\"ahler corrections in the $\mathrm{T}'$ model are less
prominent than in the \A4\ case. Our analytical treatment as well as the
\texttt{Mathematica} package (cf.\ \Secref{sec:Mathematica}) allow one to
determine the impact of these corrections in other concrete models with very
little effort.

\subsection{Further implications}
\label{sec:applications}

\subsubsection{VEV alignment}

As is well known, the VEVs of fields tend to settle at symmetry enhanced points.
However, since, as we have discussed in detail above, the full Lagrangean of
many flavor models does not really exhibit residual symmetries, one might expect
corrections also to the (holomorphic) flavon VEVs. In particular, the K\"ahler
corrections might play a role when discussing VEV alignment, i.e.\ the question
why the VEVs of the flavons take a particular form. In what follows, we make the
simplifying assumption that the flavor sector is independent of the usual
`hidden sector' which is responsible for supersymmetry breakdown. Specifically,
we assume that the $F$--term VEVs of the flavons are negligible.

Consider a model where the supersymmetric Lagrangean can be written in the form
\begin{equation}
  \mathscr{L} ~=~ 
  \left[ K(\Psi,\Psi^\dagger\, \exp{(-2 \, \mathsf{t}_A \, V_A)}) \right]_D+\left[\frac{1}{4} \, f_{AB}(\Psi) \, \mathcal{W}^A \, \mathcal{W}^B + \mathscr{W}(\Psi) + \text{h.c.} \right]_F\;,
\end{equation}
where $\Psi$ stands for all chiral superfields of the model, $V_A$ are the
vector superfields containing the gauge bosons, and $\mathcal{W}^A$ are the
corresponding field strength superfields. Then, the scalar potential, whose
minima determine the VEV structure, reads
\begin{eqnarray}
\label{eq:genScalarPotential}
  \mathscr{V}(\psi,\psi^*) & = & \left[ \frac{\partial^2 K}{\partial (\Psi^f)^\dagger \, \partial \Psi^g}(\psi,\psi^*) \right]^{-1} \cdot \frac{\partial \mathscr{W}^*}{\partial (\Psi^f)^\dagger}(\psi^*) \cdot \frac{\partial \mathscr{W}}{\partial \Psi^g}(\psi) 
  \nonumber\\
  \notag &  & {}+\frac{1}{2} \left[\re{\left(f_{AB}(\psi)\right)}\right]^{-1} \cdot \re{\left(\frac{\partial K}{\partial \Psi^f}(\psi,\psi^*) 
  \cdot (\mathsf{t}_A\, \psi)^f \right)}  \nonumber\\
  & & {}\hphantom{{}+\frac{1}{2} \left[\re{\left(f_{AB}(\psi)\right)}\right]^{-1}}\cdot
  \re{\left(\frac{\partial K}{\partial \Psi^g}(\psi,\psi^*) \cdot
  (\mathsf{t}_B\, \psi)^g \right)}\;,
\end{eqnarray}
where $\psi$ and $\psi^*$ are the scalar components of $\Psi$ and
$\Psi^\dagger$, respectively. Before taking into account the corrections to the
K\"ahler potential, the K\"ahler metric is, by assumption, diagonal,
\begin{equation}
  \frac{\partial^2 K}{\partial (\Psi^f)^\dagger \, \partial \Psi^g}(\psi,\psi^*)
  ~=~ \delta_{fg}\;,
\end{equation}
from which it follows that the scalar potential simplifies to
\begin{equation}
  \mathscr{V}(\psi,\psi^*)~=~ 
  \sum_f \left|\frac{\partial \mathscr{W}}{\partial \Psi^f}(\psi)\right|^2 + \frac{1}{2} \left[\re{\left(f_{AB}(\psi)\right)}\right]^{-1} \cdot \left(\psi^* \, \mathsf{t}_A \, \psi \right) \cdot \left(\psi^* \, \mathsf{t}_B \, \psi \right)\;.
\end{equation}
Suppose that this scalar potential has a global supersymmetric minimum at
$\psi=\psi_0$. If $\psi$ does not break supersymmetry, $\psi_0$ satisfies the
$F$--flatness and $D$--flatness conditions. 

Let us first discuss the $F$--flatness conditions. Since the K\"ahler metric is
invertible, the conditions
\begin{equation}
\label{eq:F0}
  \frac{\partial \mathscr{W}}{\partial \Psi^f}(\psi_0)~=~0,\quad \forall\, f
\end{equation}
for the case of a canonical K\"ahler potential are equivalent to the conditions
arising from \Eqref{eq:genScalarPotential}, i.e.\ for the case of an arbitrary
K\"ahler potential. This implies that K\"ahler corrections do not change the VEV
alignment via the $F$--terms.

The $D$--flatness conditions require some more care. Let us first discuss the
simplest and most common class of models, to which also the \A4 model belongs.
In these models, there is only the SM gauge symmetry, under which the flavons,
however, are not charged. Hence, the flavons do not enter the $D$--flatness
conditions irrespective of the K\"ahler potential. This, together with the
invariance of the $F$--flatness conditions, implies that the vacuum alignment is
completely untouched by the K\"ahler corrections in these models.

Let us now comment on more complicated cases. If one allows for additional
gauge symmetries such as a GUT symmetry, and also for flavons having
gauge charges, one, in principle, has to check case by case whether the VEV
alignment is changed due to modified $D$--flatness conditions. There is,
however, a simple case for which one can find a general argument. Let us assume
that the additional gauge symmetry is broken by the VEVs of one or several
chiral superfields $S^f$ which furnish irreducible representations of the gauge
group, whereas all other fields, summarized in $\Psi$ in the following, are
either not charged under the additional gauge symmetry or do not obtain a VEV.
If one can furthermore assume that the K\"ahler potential factorizes as
\begin{equation}
K\left(\Psi,\Psi^\dagger,S^f,(S^f)^\dagger \right)~=~
K_S\left((S^f)^\dagger\, \delta_{fg}\, S^g \right) \cdot K_\Psi\left(\Psi,\Psi^\dagger \right)\;,
\end{equation}
where both $K_S$ and $K_\Psi$ should contain a constant term, the $D$--flatness conditions are equivalent to the
$D$--flatness conditions of a canonical K\"ahler potential. In combination with the invariance of the $F$--flatness conditions this shows that the vacuum alignment stays completely unmodified by the K\"ahler corrections. In particular, this is fulfilled
if the gauge symmetry is only broken by the VEV of one field, i.e.\ if there is only
one field that is both charged under the gauge symmetry and attains a VEV.
This applies, for example, to the $\mathrm{T}'$ model.

In summary, we see that in most situations K\"ahler corrections will not
interfere with the usual mechanisms for VEV alignment. This, in a way, justifies to
assume that the flavons attain some `very symmetric' VEVs, as for instance in
the sample models discussed above.

\subsubsection{Constraints from FCNCs}

In supersymmetric model building, an important question concerns the flavor
structure of the soft supersymmetry breaking masses and the $A$--terms. They originate from
higher--dimensional terms in the superpotential and the K\"ahler potential. Specifically, they are induced by interactions
of the matter fields with the spurion superfield $X$ that breaks supersymmetry, i.e.\ $X \rightarrow \theta^2 \langle F_X\rangle\ne0$.

The terms relevant for our discussion are \cite{Martin:1997ns}
\begin{subequations}
\label{eq:spurionTerms}
\begin{eqnarray}
  \mathscr{W} & \supset & (Y_e)_{fg}\,L^f\,R^g\,H_d - \frac{1}{\Lambda_\mathrm{soft}}\,(Y_e^X)_{fg} \, X \,L^f\,R^g\,H_d\;, \\
  \notag K & \supset & (L^f)^\dagger \, (\mathscr{K}_L)_{fg} \, L^g
  + \frac{1}{\Lambda_\mathrm{soft}} \, \left( X \, (L^f)^\dagger \, (n_L)_{fg} \, L^g + \text{h.c.} \right)\\
  & & \qquad -~ \frac{1}{\Lambda_\mathrm{soft}^2} \, X^\dagger X \, (L^f)^\dagger \, (k_L)_{fg} \, L^g +~ L \rightarrow R\;,
\end{eqnarray}
\end{subequations}
where $\Lambda_\mathrm{soft}$ represents some messenger scale, such as the
Planck scale in the case of gravity mediation. The coupling matrices $Y_e$,
$Y^X_e$, $\mathscr{K}_{L/R}$, $n_{L/R}$ and $k_{L/R}$ are functions of the
flavon superfields and the cut--off scale $\Lambda$. All these matrices can
obtain off--diagonal entries through non--trivial flavon contractions after the
flavons acquire their VEVs. However, $\mathscr{K}_{L/R}$ and $k_{L/R}$ are
Hermitean and we choose to work in a basis where $Y_e$ is diagonal.

The supersymmetry breaking soft masses and $A$--terms can be written as
\begin{equation}
\label{eq:softTerms}
  \mathscr{L}_\mathrm{soft}~\supset~ 
  - (\widetilde{\ell}^f)^\dagger \, (\widetilde{m}^2_{\mathrm{LL}})_{fg} \, 
  \widetilde{\ell}^g
  - (\widetilde{r}^f)^\dagger \, (\widetilde{m}^2_{\mathrm{RR}})_{fg} \, 
  \widetilde{r}^g
  - \left( \widetilde{\ell}^f \, (A_{\mathrm{LR}})_{fg}\,  
  \widetilde{r}^g + \text{h.c.} \right)\;,
\end{equation}
where $\widetilde{\ell}$ and $\widetilde{r}$ denote the left-- and right--handed slepton fields, respectively.
The relations between the parameters in \Eqref{eq:spurionTerms} and \Eqref{eq:softTerms} can be obtained by replacing the spurion by its VEV and integrating out the auxiliary fields of the matter superfields. Defining $\widetilde{M}^2 = \frac{\left|\langle F_X \rangle \right|^2}{\Lambda_\mathrm{soft}^2}$, they are given by \cite{Martin:1997ns}
\begin{subequations}
\label{eq:softCoefficients}
\begin{eqnarray}
  \left(\widetilde{m}^2_{\mathrm{LL}/\mathrm{RR}}\right)_{fg}
  & = & \widetilde{M}^2 \, \left[ (k_{L/R})_{fg} 
  + (n_{L/R}^\dagger)_{fi} \, (n_{L/R})_{ig} \right]\;,\\
  (A_{\mathrm{LR}})_{fg} & = & \sqrt{\widetilde{M}^2} \, \left[ (Y_e^X)_{fg} 
  + (n_{L})_{fi} \, (Y_e)_{ig} + (Y_e)_{fi} \, (n_{R})_{ig} \right]\;. \label{eq:ATerms}
\end{eqnarray}
\end{subequations}

We now turn back to \Eqref{eq:spurionTerms} and analyze the couplings. By Schur's Lemma, the matrices  $\mathscr{K}_{L/R}$, $n_{L/R}$ and $k_{L/R}$ in the K\"ahler potential are diagonal to first order. In fact, since the left--handed lepton doublets are contained in one irreducible representation, the corresponding matrices are all proportional to the unit matrix. To simplify the following discussion, we will make the assumption that the same is true for the right--handed leptons, i.e.\
\begin{subequations}
\begin{eqnarray}
  (\mathscr{K}_{L/R})_{fg} & = & \delta_{fg}\;,\\
  (n_{L/R})_{fg} & = & \kappa_{L/R} \,\delta_{fg}\;,\\
  (k_{L/R})_{fg} & = & \kappa'_{L/R} \,\delta_{fg}\;,
\end{eqnarray}
\end{subequations}
with $\kappa_{L/R}$ and $\kappa'_{L/R}$ being order one coefficients. The restriction to generation independent coefficients does not qualitatively affect the final results.

At second order, contractions of the leptons with the flavon fields are
possible. When the flavons obtain their VEVs, the effective coupling matrices
read
\begin{subequations}
\begin{eqnarray}
  (\mathscr{K}_{L/R})_{fg} & = & \delta_{fg} - 2 \, x \, (P_{\mathrm{kin},L/R})_{fg}\;,\\
  (n_{L/R})_{fg} & = & \kappa_{L/R}\, \left[\delta_{fg} - 2 \, x \, (N_{L/R})_{fg} \right]\;,\\
  (k_{L/R})_{fg} & = & \kappa'_{L/R}\, \left[\delta_{fg} - 2 \, x \, (P_{\mathrm{soft},L/R})_{fg} \right]\;,
\end{eqnarray}
\end{subequations}
where $P_{\mathrm{kin},L/R}$ and $P_{\mathrm{soft},L/R}$ are Hermitean, $N_{L/R}$ arbitrary complex matrices, and $x$ is at most of the order of flavon VEV over the cut--off scale. In fact, most often $x$ is of the order $(\text{VEV}/\Lambda)^2$ by the arguments already outlined in \Secref{sec:linear}. We would like to emphasize that the matrices $P_{\mathrm{kin},L/R}$, $P_{\mathrm{soft},L/R}$ and $N_{L/R}$, which come from contractions of the lepton fields with the flavons, are a priori unrelated.

Since the K\"ahler metric now contains off--diagonal terms, one has to canonically normalize the lepton fields by the transformations
\begin{subequations}
\label{eq:kaehlerTrafo}
\begin{eqnarray}
  L^f & \rightarrow & L'^f = \left[\delta_{fg} + x \, (P_{\mathrm{kin},L})_{fg} \right]\,L^f\;,\\
  R^f & \rightarrow & R'^f = \left[\delta_{fg} + x \, (P_{\mathrm{kin},R})_{fg} \right]\,R^f\;,
\end{eqnarray}
\end{subequations}
which leads to the transformed coupling matrices
\begin{subequations}
\begin{eqnarray}
  (n'_{L/R})_{fg} & = & \kappa_{L/R} \, \left[\delta_{fg} + 2 \, x \left((P_{\mathrm{kin},L/R})_{fg} - (N_{L/R})_{fg} \right) \right]\;,\\
  (k'_{L/R})_{fg} & = & \kappa'_{L/R} \, \left[\delta_{fg} + 2 \, x \left((P_{\mathrm{kin},L/R})_{fg} - (P_{\mathrm{soft},L/R})_{fg} \right) \right]
\end{eqnarray}
\end{subequations}
to first order in $x$. Furthermore, one has to apply unitary transformations to the leptons that remove the off--diagonal elements from the charged lepton Yukawa matrix, in order to be able to compare the results to the experimental constraints. Since for $x=0$, i.e.\ without corrections due to the flavons, the Yukawa matrix is, by assumption, diagonal, one can write these transformations up to first order in $x$ as $U_{L/R}=1+\I \, x \, H_{L/R}$ with $H_{L/R}$ being Hermitean. Hence, this redefinition of fields does not affect $n_{L/R}$ and $k_{L/R}$ at linear order in $x$.

Hence, the soft masses for the sleptons in linear order in $x$ are
\begin{eqnarray}
  \notag (\widetilde{m}^2_{\mathrm{LL}/\mathrm{RR}})_{fg}
  & = & \widetilde{M}^2 \, \left\{ \left(\kappa'_{L/R} 
  + \left|\kappa_{L/R}\right|^2 \right) \, \delta_{fg} \right.\\
  \notag & & \quad +~ 2 \, x \, \kappa'_{L/R} \, \left[(P_{\mathrm{kin},L/R})_{fg} - (P_{\mathrm{soft},L/R})_{fg} \right]\\
  & & \quad \left. +~ 2 \, x \, \left|\kappa_{L/R}\right|^2 \,  \left[(P_{\mathrm{kin},L/R})_{fg} - (N_{L/R})_{fg} + \text{h.c.} \right] \right\}\;.
\label{eq:mLL}
\end{eqnarray}
The crucial point is that all off--diagonal terms are suppressed compared to the
diagonal terms by one factor of $x$. In special cases in which there are
relations between $P_{\mathrm{kin},L/R}$, $P_{\mathrm{soft},L/R}$ and $N_{L/R}$,
the off--diagonal terms might even vanish (almost) completely.

Before confronting this with the experimental constraints, let us first also
discuss the $A$--terms without dwelling on the details. Since after all basis
changes the Yukawa matrix $Y_e$ is diagonal, the off--diagonal elements of the
second and third term of $A_{LR}$ in \Eqref{eq:ATerms} are suppressed by one
factor of $x$. Moreover, they are suppressed by the smallness of the lepton
masses.

The coupling matrix $Y_e^X$ in \Eqref{eq:spurionTerms} can only arise from the
same flavon contractions as the Yukawa matrix $Y_e$. Neglecting the possibility
of fine--tuning, our assumption of a diagonal Yukawa thus implies diagonal
$Y_e^X$. Although the precise size of the entries of $Y_e^X$ may differ from the
lepton masses, one should assume that they are of the same order of
magnitude.\footnote{If one has a mechanism that suppresses the lepton Yukawa
couplings to the desired values, this mechanism should also suppress the entries
of $Y_e^X$ in the same way.} Since $Y_e^X$ does not have to be proportional to
$Y_e$, the effects of the transformation \eqref{eq:kaehlerTrafo} on $Y_e^X$ are
not completely undone by the unitary rotation to the charged lepton mass basis.
However, all off--diagonal terms are at most of the order $x$ and, furthermore,
suppressed by the smallness of the diagonal entries.

Let us now discuss the experimental constraints. We showed above that the
K\"ahler corrections induce off--diagonal terms for the soft masses and the
$A$--terms. Therefore, FCNCs are induced, in general by slepton, chargino,
higgsino and neutralino loops. The strongest constraints are given by the decay
$\mu \rightarrow e\,\gamma$. The SUSY contribution to this process through
photino and slepton loops is given by~\cite{Gabbiani:1996hi}
\begin{eqnarray}
 \notag \mathrm{Br}(\mu\rightarrow e\,\gamma) & = &
 \frac{12\,\pi\,\alpha^3}{G_\mathrm{F}^2\,m_\mathrm{SUSY}^4} \left|(\delta_{12})_\mathrm{LL}\,M_3(y) + \frac{\sqrt{y}\,m_\mathrm{SUSY}}{m_\mu}(\delta_{12})_\mathrm{LR}\,M_1(y) \right|^2\\
 & & \vphantom{\frac{12\,\pi\,\alpha^3}{G_\mathrm{F}^2\,m_\mathrm{SUSY}^4}}
\quad + ~ \left(\mathrm{L}\leftrightarrow\mathrm{R}\right) \;,
\label{eq:Brmutoegamma}
\end{eqnarray}
where $M_1(y)$ and $M_3(y)$ are loop factors depending on the
mass--squared ratio  between photino and slepton, $y =
m_{\widetilde{\gamma}}^{2} / m_{\widetilde{\ell}}^{2}$, where we set
$m_{\widetilde{\ell}} \approx m_\mathrm{SUSY}$.  Their precise expressions can
be found in~\cite{Gabbiani:1996hi}. For our purposes it is enough to state that 
the functions $M_{3}$ and $M_{1}$ are bounded by   $M_3 (y)  < 0.083$ and $M_1
(y) < 0.5$.  More importantly, $(\delta_{12})_\mathrm{LL}$ and
$(\delta_{12})_\mathrm{LR}$ are the mass insertion parameters, i.e.\ the ratio
between the off--diagonal and the diagonal elements of the soft masses or the
$A$--terms, respectively. Through~\Eqref{eq:mLL} we can estimate
$(\delta_{12})_\mathrm{LL}$ to be of the order of $x$. The chirality changing
mass insertion $(\delta_{12})_\mathrm{LR}$ is determined by the $A$--term, which
again is proportional to $x$. Furthermore, one might expect, as argued above,
that the $A$--term is proportional to a parameter of the order of the
corresponding Yukawa coupling. Therefore, we can estimate the $A$--term to
be of the order $\widetilde{M}\,x\,y_\mu$ , which yields
$(\delta_{12})_\mathrm{LR}~\simeq~(x\,m_\mu)/m_\mathrm{SUSY}$ for the mass
insertion parameter in \eqref{eq:Brmutoegamma}. 

In the previous sections we assumed the expansion parameter $x$ to be maximally
of the order Cabibbo angle squared, i.e.\ $x \lesssim 0.04$. Using this value in
our mass insertion parameters we can give a lower bound for the slepton mass
$m_\mathrm{SUSY}$ in order to satisfy the current experimental limit of
$\mathrm{Br}(\mu\rightarrow e\,\gamma)_{\mathrm{exp}} < 2.4 \times 10^{-12}$
\cite{Beringer:1900zz}. For a photino to slepton mass--squared ratio
of $y=5$, we have $m_\mathrm{SUSY} \geq 700 \gev$, for $y=2$, we get
$m_\mathrm{SUSY} \geq 1 \tev$,  and for $y=0.5$, we have $m_\mathrm{SUSY} \geq
1.2 \tev$. Furthermore, for $m_\mathrm{SUSY} \geq 1.4 \tev$, the experimental
limits are always satisfied, independent of the photino to slepton
mass--squared ratio. This shows that constraints from FCNCs do not
rule out sizable K\"ahler corrections for reasonable values of the soft SUSY
breaking  mass scale.

\section{Conclusions}
\label{sec:conclusions}

We have discussed the impact of K\"ahler corrections on the predictions of
models with spontaneously broken flavor symmetries. We find that these
corrections are, in general, sizable since they are controlled by the ratio of
flavon VEV over the fundamental scale, which also sets the scale 
of the expansion parameter for the entries of
the coupling and mass matrices. Furthermore, it appears hard to avoid K\"ahler
corrections because the corresponding terms cannot be forbidden by means of
conventional symmetries. In addition, the coefficients of such terms entail new
parameters, which reduce the predictivity of the respective models. In view of
these results, it appears to be premature to `rule out' certain symmetry groups
by looking at the holomorphic terms only, as has been done recently in various
scans~\cite{Parattu:2010cy,Holthausen:2012wt}.

Let us stress at this point that the situation in non--supersymmetric settings
is similar. In the non--supersymmetric case, it is, of course, 
also possible to write down higher--order corrections to the kinetic terms which
are induced by the flavon VEVs. As it turns out these induce changes of the mixing parameters identical to the supersymmetric case, therefore, extending the applicability of our discussion to non--supersymmetric models.

In particular, we have presented a full derivation of analytic formulae which
describe the change of the mixing parameters. We have applied these
formulae to two example models, one based on the flavor symmetry
$G_\mathrm{F}=\A4 \times\Z4$ \cite{Altarelli:2005yx} and one based on
$G_\mathrm{F}=\mathrm{T}' \times\Z{12}\times\Z{12}$ \cite{Chen:2009gf}. We
have demonstrated that, for the simple \A4 model which predicts
tri--bi--maximal mixing at the leading order, one of the flavon VEVs induces a
large $\theta_{13}$ value that is compatible with current experimental
limits~\cite{Abe:2011fz,An:2012eh,Ahn:2012nd}. On the other hand, the VEV
pattern in the $\mathrm{T}'$ model is such that K\"ahler corrections are not too
large unless the K\"ahler coefficients are large. This can easily be understood
with the aid of the analytic formulae derived in this paper, and can also be
checked with the associated \texttt{Mathematica} package.

Furthermore, we have shown that the K\"ahler corrections do not pose a threat to
the VEV alignment. Moreover, we have argued that they also do not induce significant
flavor changing neutral currents, i.e.\  for reasonably large soft masses, the size
of the flavor violating terms is well within the current experimental bounds.
Hence, also the vanishing of FCNCs cannot be used to constrain the K\"ahler
corrections considerably.

In conclusion, we argue that, in the supersymmetric context, a theory of flavor
requires a better understanding of the K\"ahler potential. Such an understanding
may be obtained in higher--dimensional settings, where effective couplings can
be computed from wave--function overlaps (cf.\ e.g.\
\cite{Arkani-Hamed:2001tb,Lee:2003mc}), and non--Abelian discrete symmetries may
be related to the geometry of compact space (cf.\ e.g.\
\cite{Altarelli:2006kg}). In this regard, it appears also promising to derive
flavor models from string theory, where the non--Abelian discrete symmetries
have a clear geometrical interpretation
\cite{Kobayashi:2006wq,Nilles:2012cy,BerasaluceGonzalez:2012vb}. In certain
settings, the K\"ahler potentials are known to some extent \cite{Cvetic:1988yw};
some information can be inferred from the transformation behavior of the fields
under the modular group \cite{Dixon:1989fj,Ibanez:1992hc}; however, closed
expressions for higher--order terms have not yet been worked out.

\subsection*{Acknowledgments}

We would like to thank C.~Albright and J.~Heckman for useful discussions. 
M.-C.C.\ would like to thank TU M\"unchen, where part of the work was done, for
hospitality. M.R.\ would like to thank the UC Irvine, where part of this work was
done, for hospitality. This work was partially supported by the Deutsche
Forschungsgemeinschaft (DFG) through the cluster of excellence ``Origin and
Structure of the Universe'' and the Graduiertenkolleg ``Particle Physics at the
Energy Frontier of New Phenomena''. This research was done in the context of the
ERC Advanced Grant project ``FLAVOUR''~(267104), and was partially supported by
the U.S. National Science Foundation under Grant No.\ PHY-0970173. We thank the
Aspen Center for Physics, where this discussion was initiated,  the Galileo
Galilei Institute for Theoretical Physics (GGI),  the Simons Center for Geometry
and Physics in Stony Brook,  and the Center for Theoretical Underground Physics
and Related Areas (CETUP* 2012) in South Dakota for their hospitality and for
partial support during the completion of this work.


\appendix

\section{Conventions}

\subsection{Parametrization of \boldmath $U_\mathrm{PMNS}$ \unboldmath}
\label{app:parametrization}
The parametrization of the PMNS--matrix used in this text is shown here. First, $U_\mathrm{PMNS}$ is decomposed in the product of a diagonal phase matrix containing the unphysical lepton phases, a CKM--like matrix $V$ and a diagonal matrix containing the two Majorana phases,
\begin{equation}
  U_\mathrm{PMNS}~=~\diag{(\mathrm{e}^{\I \delta_e},\,\mathrm{e}^{\I \delta_\mu},\,\mathrm{e}^{\I \delta_\tau})} \cdot V(\theta_{12},\,\theta_{13},\,\theta_{13},\,\delta) \cdot \diag{(\mathrm{e}^{-\I \phi_1/2},\,\mathrm{e}^{\I \phi_2/2},\,1)}\;.
\end{equation}
The matrix $V$ itself is parametrized as
\begin{equation}
  V~=~
  \begin{pmatrix}
    c_{12} \, c_{13} & s_{12} \, c_{13} & s_{13} \, \mathrm{e}^{- \I \delta}\\
    - s_{12} \, c_{23} - c_{12}\, s_{23}\, s_{13}\, \mathrm{e}^{\I \delta} & c_{12}\, c_{23} - s_{12}\, s_{23}\, s_{13}\, \mathrm{e}^{\I \delta} & s_{23}\, c_{13}\\
    s_{12}\, s_{23} - c_{12}\, c_{23}\, s_{13}\, \mathrm{e}^{\I \delta} & - c_{12}\, s_{23} - s_{12}\, c_{23}\, s_{13}\, \mathrm{e}^{\I \delta} & c_{23}\, c_{13}
  \end{pmatrix}\;.
\end{equation}
Here, $s_{ij}$ denotes $\sin \theta_{ij}$ and $c_{ij}$ denotes $\cos \theta_{ij}$.

\subsection{\boldmath $\A4$ \unboldmath}
\label{app:A4}
In section \ref{sec:Kaehler} we provided possible K\"ahler corrections for models based on the non--Abelian flavor group $\A4$. In this appendix, we recall the most important aspects of the $\A4$ group, which is the symmetry group of the regular tetrahedron. It has four inequivalent irreducible representations, including three singlets $\rep{1}\,, \rep{1}'\,, \rep{1}''$ and one triplet $\rep{3}$. Throughout the literature there are mainly two different bases that have been used for $\A4$. In \Secref{sec:Kaehler} we utilize the basis in which the generators $S$ and $T$ are represented as 
\begin{equation}
 S=\frac{1}{3}\left(\begin{array}{ccc}
                       -1 & 2 & 2 \\ 2 & -1 & 2 \\ 2 &2 & -1
                      \end{array}\right)\;, \qquad
T=\left(\begin{array}{ccc}
                       1 & 0 & 0 \\ 0 & \omega^2 & 0 \\ 0 & 0 & \omega
                      \end{array}\right)\;, \qquad 
\mathrm{with} \qquad \omega=\mathrm{e}^{\frac{2 \pi \I}{3}}\;.
\end{equation}
These generators give us the multiplication rule
\begin{equation}\label{eq:3x3}
 \rep{3}\otimes\rep{3}~=~\rep{1} \oplus \rep{1'} \oplus \rep{1''} \oplus \rep{3}_\mathrm{s} \oplus \rep{3}_\mathrm{a}\;,
\end{equation}
where $\rep{3}_\mathrm{s}$ and $\rep{3}_\mathrm{a}$ denote the symmetric and
the antisymmetric triplet combinations, respectively.
In terms of the components of the two triplets, $\rep{a}$ and $\rep{b}$,
\begin{subequations}
\label{eq:A4multi}
\begin{eqnarray}
 \left(\rep{a} \otimes \rep{b} \right)_{\rep{1}}
 & = & 
 a_{1}\,b_{1}+a_{2}\,b_{3}+a_{3}\,b_{2}\;,
 \\ 
 \left(\rep{a}\otimes \rep{b} \right)_{\rep{1'}}
 & = & 
 a_{2}\,b_{2}+a_{1}\,b_{3}+a_{3}\,b_{1}\;, \\
 \left(\rep{a} \otimes \rep{b} \right)_{\rep{1''}}
 & = & 
 a_{3}\,b_{3}+a_{1}\,b_{2}+a_{2}\,b_{1}\;, \\
\left(\rep{a} \otimes \rep{b} \right)_{\rep{3}_\mathrm{s}}
 & = & \frac{1}{\sqrt{2}}\,
\left(\begin{array}{c}
 2 a_{1}\,b_{1} - a_{2}\,b_{3} - a_{3}\,b_{2}\\
 2 a_{3}\,b_{3} - a_{1}\,b_{2} - a_{2}\,b_{1}\\
 2 a_{2}\,b_{2} - a_{1}\,b_{3} - a_{3}\,b_{1}
\end{array}\right)\;, 
\\
 \left( \rep{a} \otimes \rep{b} \right)_{\rep{3}_\mathrm{a}}
 & = & \I\,\sqrt{\frac{3}{2}}\,
\left(\begin{array}{c}
 a_{2}\,b_{3} - a_{3}\,b_{2}\\
 a_{1}\,b_{2} - a_{2}\,b_{1}\\
 a_{3}\,b_{1} - a_{1}\,b_{3}
\end{array}\right)\;, 
\end{eqnarray}
\end{subequations}
where $\left(\rep{a} \otimes \rep{b} \right)_{\rep{R}}$ indicates that $\rep{a}$ and $\rep{b}$ are contracted to 
the representation \rep{R}. Note that there are different conventions for
normalizing the triplets $\rep{3}_i$ in the literature, and
the corresponding factors can be absorbed in the K\"ahler coefficients.

In another basis, $\A4$ is generated by 
\begin{equation}
 \widetilde{S}~=~\left(\begin{array}{ccc}
                       1 & 0 & 0 \\ 0 & -1 & 0 \\ 0 & 0 & -1
                      \end{array}\right)\;, \qquad
\widetilde{T}~=~\left(\begin{array}{ccc}
                       0 & 0 & 1 \\ 1 & 0 & 0 \\ 0 & 1 & 0
                      \end{array}\right)\;,
\end{equation}
which is related to our basis through the unitary transformation matrix 
\begin{equation}
 U_{\omega} ~=~ \frac{1}{\sqrt{3}}\left(\begin{array}{ccc}
1 & 1 & 1 \\
1 & \omega & \omega^2 \\
1 & \omega^2 & \omega
\end{array}\right)\;.
\end{equation}
The relation between the two bases is then given by $\widetilde{S}=U_{\omega}\,S\,U_{\omega}^{\dagger}$ and $\widetilde{T}=U_{\omega}\,T\,U_{\omega}^{\dagger}$.

It is important to note that this basis transformation also relates the different flavon VEVs to one another. This means that the VEV $(v,v,v)^T$ in one basis is equivalent to the VEV $(v',0,0)^T$ in the other basis, and vice versa.

\section{Examples for analytic formulae}
\label{app:ExForm}

We present examples of the analytic formulae for corrections due to
$P_L$ in 
\begin{equation}
  K~=~L^\dagger\, (1+x_L\,P_L)\, L + R^\dagger\, R\;,
\end{equation}
where $P_L$ is replaced by one of the nine basis matrices $P_i$, as shown in 
\Eqref{eq:Pmatrices}. We take the tri--bi--maximal mixing as initial condition for
the mixing parameters, i.e.\
\begin{eqnarray}
  \theta_{12} & = & \arcsin{\frac{1}{\sqrt{3}}}\;, \quad 
  \theta_{13}~=~0,\quad 
  \theta_{23}~=~\frac{\pi}{4}\;,\quad 
  \delta~=~\text{undefined}\;,  \nonumber \\ 
  \delta_e & = & \pi\;,\quad 
  \delta_\mu~=~\pi\;,\quad  
  \delta_\tau~=~0\;,\quad 
  \varphi_1~=~\varphi_2~=~2\pi\;.
\end{eqnarray}
The CP phase $\delta$ is determined from the formulae by demanding that the
change of $\delta$ is analytical at $\theta_{13}=0$ for each of the $P_i$, which
yields $\delta=0$ for $i=1,\dots,6$ and $\delta=-\pi/2$ for $i=7,8,9$. The
neutrino masses $m_i$ are left unspecified. The pronounced hierarchy of the
charged lepton masses, i.e.\ $m_\tau \gg m_\mu \gg m_e$,  is used to simplify
the results. In leading order in an
expansion in the small mass ratios, the charged lepton masses completely
drop out from the formulae. We obtain the following analytical expressions
for the changes of the mixing angles:
\begin{itemize}
\item For $P=P_1$:
\begin{subequations}
\begin{eqnarray}
  \Delta \theta_{12}^{(1)} & = & x_L\,\frac{1}{3\,\sqrt{2}}\, \frac{m_1+m_2}{m_1-m_2}\;,\\
  \Delta \theta_{13}^{(1)} & = & 0\;,\\
  \Delta \theta_{23}^{(1)} & = & 0\;.
\end{eqnarray}
\end{subequations}

\item For $P=P_2$:
\begin{subequations}
\begin{eqnarray}
  \Delta \theta_{12}^{(2)} & = & x_L\,\frac{1}{3\,\sqrt{2}}\, \frac{2\,m_1-m_2}{m_1-m_2}\;,\\
  \Delta \theta_{13}^{(2)} & = & x_L\,\frac{1}{3\,\sqrt{2}}\, \frac{3\,m_1\,m_2-2\,m_1\,m_3-m_2\,m_3}{(m_1-m_3)\,(m_2-m_3)}\;,\\
  \Delta \theta_{23}^{(2)} & = & x_L\,\frac{1}{3}\, \frac{m_3\,(m_1-m_2)}{(m_1-m_2)\,(m_2-m_3)}\;.
\end{eqnarray}
\end{subequations}

\item For $P=P_3$:
\begin{subequations}
\begin{eqnarray}
  \Delta \theta_{12}^{(3)} & = & -x_L\,\frac{1}{6\,\sqrt{2}}\, \frac{m_1+m_2}{m_1-m_2}\;,\\
  \Delta \theta_{13}^{(3)} & = & x_L\,\frac{1}{3\,\sqrt{2}}\, \frac{m_3\,(m_1-m_2)}{(m_1-m_3)\,(m_2-m_3)}\;,\\
  \Delta \theta_{23}^{(3)} & = & x_L\,\frac{1}{12}\, \frac{m_1\,(3\,m_2+m_3)-m_3\,(m_2+3\,m_3)}{(m_1-m_3)\,(m_2-m_3)}\;.
\end{eqnarray}
\end{subequations}

\item For $P=P_4$:
\begin{subequations}
\begin{eqnarray}
  \Delta \theta_{12}^{(4)} & = & x_L\,\frac{1}{3\,\sqrt{2}}\, \frac{2\,m_1-m_2}{m_1-m_2}\;,\\
  \Delta \theta_{13}^{(4)} & = & -x_L\,\frac{1}{3\,\sqrt{2}}\, \frac{3\,m_1\,m_2-2\,m_1\,m_3-m_2\,m_3}{(m_1-m_3)\,(m_2-m_3)}\;,\\
  \Delta \theta_{23}^{(4)} & = & -x_L\,\frac{1}{3}\, \frac{m_3\,(m_1-m_2)}{(m_1-m_3)\,(m_2-m_3)}\;.
\end{eqnarray}
\end{subequations}

\item For $P=P_5$:
\begin{subequations}
\begin{eqnarray}
 \Delta \theta_{12}^{(5)} & = & -x_L\,\frac{1}{3\,\sqrt{2}}\, \frac{m_1 + m_2}{m_1 - m_2}\;, \\
 \Delta \theta_{13}^{(5)} & = & 0\;, \\
 \Delta \theta_{23}^{(5)} & = & - x_L\, \frac{1}{2}\;.
\end{eqnarray}
\end{subequations}

\item For $P=P_6$:
\begin{subequations}
\begin{eqnarray}
 \Delta \theta_{12}^{(6)} & = & -x_L\,\frac{1}{6\,\sqrt{2}}\, \frac{m_1 + m_2}{m_1 - m_2}\;, \\
 \Delta \theta_{13}^{(6)} & = & -x_L\,\frac{1}{3\,\sqrt{2}}\, \frac{m_3\, (m_1 - m_2)}{(m_1 - m_3)\,(m_2-m_3)}\;, \\
 \Delta \theta_{23}^{(6)} & = & x_L\,\frac{1}{12}\, \frac{m_1\,(3\,m_2 + m_3)-m_3\,(m_2+3\,m_3)}{(m_1 - m_3)\,(m_3-m_2)}\;.
\end{eqnarray}
\end{subequations}

\item For $P=P_7$:
\begin{subequations}
\begin{eqnarray}
 \Delta \theta_{12}^{(7)} & = & 0\;, \\
 \Delta \theta_{13}^{(7)} & = & -x_L\,\frac{1}{3\,\sqrt{2}}\, \frac{3\,m_1\,m_2 + 2\,m_1\,m_3 + m_2\,m_3}{(m_1+m_3)\,(m_2+m_3)}\;, \\
 \Delta \theta_{23}^{(7)} & = & 0\;.
\end{eqnarray}
\end{subequations}

\item For $P=P_8$:
\begin{subequations}
\begin{eqnarray}
 \Delta \theta_{12}^{(8)} & = & 0\;, \\
 \Delta \theta_{13}^{(8)} & = & x_L\,\frac{1}{3\,\sqrt{2}}\, \frac{3\,m_1\,m_2 + 2\,m_1\,m_3 + m_2\,m_3}{(m_1+m_3)\,(m_2+m_3)}\;, \\
 \Delta \theta_{23}^{(8)} & = & 0\;.
\end{eqnarray}
\end{subequations}

\item For $P=P_9$:
\begin{subequations}
\begin{eqnarray}
 \Delta \theta_{12}^{(9)} & = & 0\;, \\
 \Delta \theta_{13}^{(9)} & = & -x_L\,\frac{\sqrt{2}}{3}\, \frac{m_3\,(m_1-m_2)}{(m_1 + m_3)\,(m_2 + m_3)}\;, \\
 \Delta \theta_{23}^{(9)} & = & 0\;.
\end{eqnarray}
\end{subequations}
\end{itemize}
As discussed in the main text, a general $P$ matrix can be decomposed into
the nine basis matrices $P_i$, 
\begin{equation}
 P~=~\sum\limits_{i=1}^9 x_L^{(i)}\,P_i\;,
\end{equation}
and the resulting changes for the mixing angles are then given by
\begin{equation}
 \Delta\theta_{jk}~=~\sum\limits_{i=1}^9 x_L^{(i)}\,\Delta\theta_{jk}^{(i)}\;.
\end{equation}
With our \texttt{Mathematica} package (cf.\ \Secref{sec:Mathematica}) one can derive similar expressions for
other initial conditions on the mixing parameters.

\bibliography{Orbifold}
\addcontentsline{toc}{section}{Bibliography}
\bibliographystyle{NewArXiv} 
\end{document}